\documentclass[aps,10pt,prd,twocolumn,superscriptaddress,preprintnumbers,%
showkeys,nofootinbib,showpacs]{revtex4-1}
\usepackage{bm}
\usepackage{graphicx}
\usepackage[hypertexnames=false]{hyperref}
\usepackage{amsmath,amssymb,amsfonts,latexsym}
\newcommand{\be}{\begin{eqnarray}}
\newcommand{\ee}{\end{eqnarray}}
\newcommand{\beq}{\begin{equation}}
\newcommand{\eeq}{\end{equation}}
\begin{document}
\title{Transverse charge and current densities in the nucleon \\ from dispersively improved
chiral effective field theory}
\author{J.~M.~Alarc\'on}
\email[ E-mail: ]{jmanuel.alarcon@uah.es}
\affiliation{University of Alcal\'a, Nuclear and Particle Physics Group,
Department of Physics and Mathematics, 28805 Alcal\'a de Henares (Madrid), Spain}
\author{C.~Weiss}
\email[ E-mail: ]{weiss@jlab.org}
\affiliation{Theory Center, Jefferson Lab, Newport News, VA 23606, USA}
\begin{abstract}
\begin{description}
\item[Background] The transverse densities $\rho_{1, 2}(b)$ describe the distributions of electric charge
and magnetic moment at fixed light-front time and connect the nucleon's elastic form factors
with its partonic structure. The dispersive representation of the form factors $F_{1, 2}(t)$
expresses the densities in terms of exchanges of hadronic states in the $t$-channel and permits
their analysis using hadronic physics methods.
\item[Purpose] Compute the densities at peripheral distances $b = \mathcal{O}(M_\pi^{-1})$, where they
are generated predominantly by the two-pion states in the dispersive representation. Quantify the uncertainties.
\item[Methods]
Dispersively improved chiral effective field theory (DI$\chi$EFT) is used to calculate
the isovector spectral functions $\textrm{Im}\, F_{1, 2}(t)$ on the two-pion cut. The method includes
$\pi\pi$ interactions ($\rho$ resonance) through elastic unitarity and provides realistic spectral
functions up to $t \approx$ 1 GeV$^2$. Higher-mass states are parametrized by effective
poles and constrained by sum rules (charges, radii, superconvergence relations). The densities
$\rho_{1, 2}(b)$ are obtained from their dispersive representation. Uncertainties are quantified
by varying the spectral functions. The method respects analyticity and ensures
the correct $b \rightarrow \infty$ asymptotic behavior of the densities.
\item[Results] Accurate densities are obtained at all distances $b \gtrsim 0.5$ fm,
with correct behavior down to $b \rightarrow 0$. The region of distances is quantified
where transverse nucleon structure is governed by the two-pion state.
The light-front current distributions in the polarized nucleon are computed and discussed.
\item[Conclusions] Peripheral nucleon structure can be computed from first principles using DI$\chi$EFT.
The method can be extended to generalized parton distributions and other nucleon form factors.
\end{description}
\end{abstract}
\maketitle
\tableofcontents
\section{Introduction}
Transverse densities have emerged as a key concept in nucleon structure physics.
The functions $\rho_{1, 2} (b)$ describe the transverse coordinate distributions of charge
and current in the nucleon at fixed light-front time $x^+ = x^0 + x^3$
and provide a spatial representation appropriate to the relativistic nature of the dynamical
system \cite{Soper:1976jc,Burkardt:2000za,Burkardt:2002hr,Miller:2007uy}.
They are defined as two-dimensional Fourier transforms of the invariant
form factors (FFs) $F_{1, 2}(t)$ parametrizing the current matrix element between nucleon states.
At the same time, they represent a projection of the generalized
parton distributions (GPDs) describing the distribution of partons in light-front
longitudinal momentum and transverse position \cite{Burkardt:2000za,Burkardt:2002hr}.
As such, the transverse densities connect the FFs measured in low-energy 
electron-nucleon elastic scattering with the partonic structure probed in high-energy
processes such as deep-inelastic scattering and hard exclusive processes.

The nucleon FFs $F_{1, 2}(t)$ at spacelike momentum transfers $t < 0$ can be interpreted
in terms of hadronic exchanges between the current and the nucleon. The mathematical framework
is provided by the dispersive representation of the FFs based on analyticity in $t$.
The FFs at $t < 0$ are expressed as integrals over their imaginary parts on the cut at $t > t_{\rm thr} > 0$,
$\textrm{Im}\, F_{1, 2}(t)$,
the so-called spectral functions, which correspond to $t$-channel states with definite hadronic composition
and quantum numbers. A similar dispersive representation can be derived for the
transverse densities \cite{Strikman:2010pu}. It establishes a correspondence between the
densities at a given distance $b$ and the hadronic exchanges at various masses
$t > 0$ \cite{Miller:2011du}. In particular, it connects the densities at large distances
$b \gtrsim$ 1 fm with the lowest-mass $t$-channel states and permits a systematic study
of ``peripheral'' nucleon structure \cite{Strikman:2010pu,Granados:2013moa,Granados:2015rra,Granados:2016jjl}.
Using this framework one can compute the peripheral densities from first principles employing methods
of hadronic physics. One can also explore the duality between the hadronic exchanges in the $t$-channel
and the partonic structure in $s$-channel \cite{Miller:2011du}.

The lowest-mass $t$-channel state in the nucleon electromagnetic FFs is the two-pion ($\pi\pi$) state.
It appears in the isovector channel and saturates the isovector spectral functions up to
$t \approx$ 1 GeV$^2$. The $\pi\pi$ system in this mass region interacts strongly and forms the
$\rho$ resonance at $t \approx 0.6$ GeV$^2$. The picture of ``vector dominance'' abstracted from
this situation explains many observations in the phenomenology of the electromagnetic form factors
of the nucleon and other hadrons. The spectral functions in the $\pi\pi$ channel have been
constructed empirically using methods of hadronic amplitude analysis, such as elastic unitarity
with input from $\pi N$ scattering data \cite{Hohler:1976ax,Belushkin:2005ds}, or Roy-Steiner
equations \cite{Hoferichter:2016duk}. The transverse densities have been studied using the
dispersive representation with such empirical spectral functions \cite{Miller:2011du}.

It would be interesting if the transverse densities could be computed using spectral functions
derived from chiral effective field theory (chiral EFT). This approach would open up several new possibilities.
First, chiral EFT is predictive and allows one to reduce the information content of the spectral
functions and densities to a few universal parameters (low-energy constants), which are determined
from independent measurements. Second, one can quantify the uncertainties in the peripheral densities
using the parametric expansion of chiral EFT. Third, because chiral EFT is a point-particle field theory,
one can explore the duality between $t$-channel exchanges and $s$-channel structure at a
microscopic level and derive a partonic representation of the chiral processes.
Fourth, because of the universality of chiral EFT one can relate the electromagnetic densities
to other elements of peripheral nucleon structure.

Traditional chiral EFT calculations of the spectral functions are limited to the near-threshold
region $t - 4 M_\pi^2 \sim$ few $M_\pi^2$ and cannot describe the $\rho$ resonance region,
because the strong $\pi\pi$ interactions amount to large higher-order
corrections \cite{Gasser:1987rb,Bernard:1996cc,Kubis:2000zd,Kaiser:2003qp}.
When applied to the transverse densities, this only allows one to compute the densities
at very large distances $b \gtrsim$ 2 fm, where they are extremely small and of little
practical interest \cite{Strikman:2010pu,Granados:2013moa,Granados:2015rra,Granados:2016jjl}.
In order to go to smaller distances, one needs an approach that takes into account the $\pi\pi$
interactions in the $\rho$ region in a different manner. Dispersively improved chiral EFT (DI$\chi$EFT)
is an approach that incorporates $\pi\pi$ interactions through elastic unitarity and enables
EFT-based calculations of the spectral functions in the $\rho$ meson
region \cite{Alarcon:2017ivh,Alarcon:2017lhg,Alarcon:2018irp}
(an equivalent alternative formulation is described in Ref.~\cite{Granados:2017cib}).
Using the $N/D$ method,
the spectral function is separated into a part containing the nonperturbative interactions in the $\pi\pi$ system,
which is taken from the measured pion timelike form factor, and a part describing the coupling
of the $\pi\pi$ system to the nucleon, which can be computed with chiral EFT with good convergence.
The DI$\chi$EFT spectral functions have been used to compute the electromagnetic form factors
\cite{Alarcon:2017lhg,Alarcon:2018irp}
and extract the proton radii from elastic scattering data \cite{Alarcon:2018zbz,Alarcon:2020kcz};
the approach has also been applied to the nucleon scalar FF \cite{Alarcon:2017ivh}.
A first study of transverse densities has been performed in leading-order (LO) accuracy \cite{Alarcon:2017asr}.

In this work we use DI$\chi$EFT to compute the peripheral transverse charge and magnetization
densities $\rho_{1, 2}(b)$ in the nucleon and study their properties. We construct the $\pi\pi$
spectral functions $\textrm{Im}\, F_{1, 2}(t)$ in partial next-to-leading-order (N2LO) accuracy.
High-mass states are described by effective poles, whose parameters are fixed by dispersive sum rules and
superconvergence relations. We compute the transverse densities in the dispersive representation
and quantify the region of distances where they are dominated by the $\pi\pi$ state.
We quantify the uncertainties of the densities resulting from the low-energy constants and the high-mass poles.
We also compute the transverse light-front current densities and construct 2-dimensional images
of the transversely polarized nucleon. We discuss the interpretation of the results and
possible applications to physics studies with transverse densities.

Novel aspects of the present study of transverse densities are as follows:
(a)~The dispersive representation respects the analytic properties of the FFs
(position and strength of singularities) and produces densities with the correct asymptotic
behavior at $b \rightarrow \infty$. This makes it possible to reliably compute the peripheral
densities and estimate their uncertainties. Methods based on the Fourier transform of empirical
FFs become unstable at large $b$ and are not adequate for peripheral densities \cite{Venkat:2010by}.
(b)~The DI$\chi$EFT approach incorporates $\pi\pi$ interactions and the $\rho$ resonance and
produces realistic spectral functions up to $t \approx 1$ GeV$^2$, which allows one to
compute the densities down to distances $b \lesssim 0.5$ fm, substantially smaller than
possible with traditional chiral EFT \cite{Strikman:2010pu,Granados:2013moa,Granados:2015rra,Granados:2016jjl}.
This means that a large fraction
of transverse nucleon structure is now amenable to an EFT description and can be deconstructed
in terms of effective degrees of freedom and low-energy constants, representing a significant
gain of information. It also means that the EFT description at large distances can be matched with
a quark model-based description of the densities at distances $b \lesssim 1$ fm,
enabling studies of quark-hadron duality in the transverse densities.
(c)~The uncertainties of the densities are estimated in the context of the dispersive representation,
by varying the elements of the spectral functions. The unknown high-mass part of the spectral functions
is parametrized by a random ensemble of high-mass poles, whose distribution is constrained by
stability criteria imposed on the spacelike form factors. This new formulation minimizes the model
dependence in the description of the high-mass states \cite{Alarcon:2017lhg,Alarcon:2018irp}
and enables robust uncertainty estimates for the densities.

The article is organized as follows. In Sec.~\ref{sec:methods} we describe the methods used in the
present study, including the properties of the transverse densities, their dispersive representation,
the construction of the DI$\chi$EFT spectral functions, and the uncertainty estimates in the
dispersive representation. In Sec.~\ref{sec:results} we describe the results, including the
DI$\chi$EFT isovector spectral functions, the isovector transverse densities,
the proton and neutron densities, and the light-front current densities in the polarized nucleon.
In Sec.~\ref{sec:discussion} we discuss the results and outline possible future applications to
quark-hadron duality and other structures. In Appendix~\ref{app:radii} we collect the nucleon
radii and their uncertainties, which serve as input parameters in the DI$\chi$EFT calculation.
In Appendix~\ref{app:spectral} we summarize the formulas for the $N$ functions appearing
in the calculation of the spectral functions with the $N/D$ method.
In Appendix~\ref{app:isoscalar} we describe the parametrization of the isoscalar spectral
functions used in the calculation of proton and neutron densities.
\section{Methods}
\label{sec:methods}
\subsection{Transverse densities}
The transition matrix element of the electromagnetic current operator between nucleon (proton, neutron)
states with 4-momentum transfer $\Delta \equiv p' - p$ is described by the FFs $F_1(t)$ and $F_2(t)$
(Dirac and Pauli FFs); see Ref.~\cite{Alarcon:2017asr} for details. They are functions of the
invariant momentum transfer $t = \Delta^2$, with $t < 0$ in the physical region of elastic scattering.
Their values at $t = 0$ are given by the nucleon charges and anomalous magnetic moments
(in units of nuclear magnetons),
\begin{align}
F_1^{p, n} (0) \; &= \; Q^{p, n} \; = \; (1, 0),
\\
F_2^{p, n} (0) \; &= \; \kappa^{p, n}  \; = \; (1.793, -1.913).
\end{align}
The isovector and isoscalar components are defined as
\be
F_i^{V, S} \; \equiv \; {\textstyle\frac{1}{2}}(F_i^p \mp F_i^n) \hspace{2em} (i = 1, 2);
\ee
the same convention is used for other quantities (densities, radii; see below).

The FFs are invariant functions and can be analyzed without specifying the form of relativistic
dynamics or choosing a reference frame. Their interpretation in terms of spatial distributions
requires specific choices. In the light-front form of relativistic dynamics one follows the
evolution of strong interactions in light-front time $x^+ \equiv x^0 + x^3$ and describes
the structure of systems at fixed $x^+$ \cite{Dirac:1949cp,Leutwyler:1977vy,Lepage:1980fj}.
In this context it is natural to consider the FFs
in a class of reference frames where the 4-momentum transfer has only transverse components,
$\Delta^0 = \Delta^3 = 0, \bm{\Delta}_T \equiv (\Delta^1, \Delta^2) \neq 0$, with $|\bm{\Delta}_T|^2 = -t$.
The FFs can then be represented as 2-dimensional Fourier integrals over a transverse coordinate
variable $\bm{b}$, with $b \equiv |\bm{b}|$,
\begin{align}
F_i (t = -|\bm{\Delta}_T|^2) &= \int d^2 b \; 
e^{i \bm{\Delta}_T \cdot \bm{b}} \; \rho_i (b)
\hspace{2em} (i = 1, 2)
\label{rho12_Fourier}
\end{align}
(the two-dimensional Fourier integrals can also be expressed as radial Fourier-Bessel integrals)
\cite{Soper:1976jc,Burkardt:2000za,Burkardt:2002hr,Miller:2007uy}.
The functions $\rho_{1, 2}(b)$ are the transverse densities. Their spatial integrals
reproduce the total charge and anomalous magnetic moment,
\begin{align}
\int d^2 b \; \rho_1^{p, n} (b) &= F_1^{p, n}(0) = Q^{p, n} ,
\\
\int d^2 b \; \rho_2^{p, n} (b) &= F_2^{p, n}(0) = \kappa^{p, n} .
\label{rho_2_integral}
\end{align}
The functions $\rho_{1, 2}(b)$ describe the transverse spatial distributions of electric
charge and anomalous magnetic moment in the nucleon at fixed light-front time.
The distributions are frame-independent (they are invariant under longitudinal
light-front boosts and transform kinematically under transverse boosts) and
provide a spatial representation appropriate to the relativistic nature of the dynamical
system; see Refs.~\cite{Burkardt:2000za,Miller:2010nz,Granados:2013moa} for discussion.

%
%
\begin{figure}
\includegraphics[width=.26\textwidth]{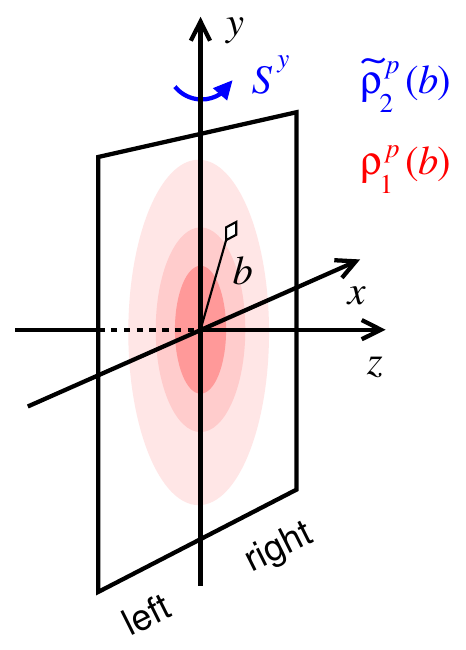}
\caption[]{Interpretation of the transverse densities in a proton state
localized at the transverse origin, $x = y = 0$, Eq.~(\ref{j_plus_rho}).
$\rho_1(b)$ describes the spin-independent $J^+$ current
at transverse radius $b$. $\widetilde \rho_2(b)$ describes the spin-dependent
distortion in a nucleon polarized in the $y$-direction.}
\label{fig:rho12t_concept}
\end{figure}
A simple interpretation of the densities can be provided in nucleon states which
are localized in transverse position (see Fig.~\ref{fig:rho12t_concept};
we use $x, y, z$ to denote the $1, 2, 3$ spatial directions) \cite{Burkardt:2000za,Granados:2013moa}.
In a state where the nucleon is localized at the transverse origin,
and its spin quantized along the $y$-direction, the expectation value
of the current $J^+$ at the transverse position $\bm{b}$ is
\begin{align}
\langle J^+ (\bm{b}) \rangle^p_{\textrm{\scriptsize localized}}
& = (...) \left[\rho_1^p (b) + (2 S^y) \cos\phi \, \widetilde\rho_2^{\, p} (b) \right] , 
\label{j_plus_rho}
\\[2ex]
\widetilde\rho_2^{\, p} (b) & \equiv \frac{\partial}{\partial b} 
\left[ \frac{\rho_2^p (b)}{2 m_N} \right] 
\label{rho_2_tilde_def}
\end{align}
(same for $p \rightarrow n$). Here $(...)$ represents a factor resulting from the normalization
of states \cite{Granados:2013moa}, $\phi$ is the angle of the vector $\bm{b}$ relative to the $x$--axis,
and $S^y = \pm 1/2$ is the expectation value of the spin in the $y$--direction.
$m_N$ is the nucleon mass (same for $p$ and $n$). Thus $\rho_1(b)$ describes the spin-independent
part of the plus current in the localized nucleon state, and $\widetilde\rho_2 (b)$ describes
the spin- and angle-dependent part of the current in a transversely polarized nucleon.
Note that $\widetilde\rho_2 (b)$ satisfies the integral relation
\begin{align}
\int d^2 b \; b \, \widetilde \rho_2^{\, p, n} (b) &= -\frac{\kappa^{\, p, n}}{m_N} ,
\end{align}
which is obtained from Eq.(\ref{rho_2_integral}) by integration by parts.

The derivatives of the FFs at $t = 0$ are related to the average squared
transverse radii of the distributions,
\begin{align}
4 \frac{dF_1^{p, n}}{dt}(0) &= \langle b^2 \rangle_1^{p, n}
\equiv \int d^2b \, b^2 \, \rho_1^{p, n}(b),
\\[1ex]
\frac{4}{\kappa^{p, n}} \frac{dF_2^p}{dt}(0) &= \langle b^2 \rangle_2^{p, n}
\equiv \frac{1}{\kappa^{p, n}} \int d^2b \, b^2 \, \rho_2^{p, n}(b) .
\end{align}
The factor 4 results from the 2-dimensional distributions
and replaces the well-known factor 6 in the representation of the FF derivatives in terms
of conventional 3-dimensional radii. While the 3-dimensional radii have a physical interpretation
only in nonrelativistic systems, the transverse radii here are averages of spatial distributions
that have a well-defined meaning for relativistic system. The relation between the transverse radii
and the 3-dimensional Dirac and Pauli radii 
$\langle r^2 \rangle_{1, 2}$ is
\begin{align}
\langle b^2 \rangle_1^{p, n} = \frac{2}{3} \langle r^2 \rangle_1^{p, n}
\hspace{2em}
\langle b^2 \rangle_2^{p, n} =
\frac{2}{3}
\langle r^2 \rangle_2^{p, n} .
\label{b2_r2}
\end{align}
The nucleon radii are used as parameters in the dynamical
calculations of the transverse densities in this work. Because form factor phenomenology
usually quotes the 3-dimensional nucleon radii, we present our calculations such
that they use the 3-dimensional radii as input, keeping in mind that they are
related to the transverse radii by Eq.~(\ref{b2_r2}). The empirical values of the
3-dimensional radii and their uncertainties are summarized in Appendix~\ref{app:radii}
and will be quoted in the following.

In studies of the nucleon's partonic structure in QCD one considers the transverse
coordinate distributions of quarks and antiquarks with a given light-cone momentum
fraction $x$ in the proton, $f^a (x, b)$ and $\bar f^a (x, b)$, where $a = u, d, ...$
denotes the quark flavor. They are defined as the Fourier representation of the GPDs
$H^a(x, \xi = 0, t)$, which describe the form factors of partons with light-cone plus
momentum fraction $x$ in the proton, in the situation where the plus momentum difference
between the proton states is $\xi = 0$ and the momentum transfer has
only transverse components \cite{Burkardt:2000za,Burkardt:2002hr},
\begin{align}
H^a (x, \xi=0, t = -|\bm{\Delta}_T|^2) &= \int d^2 b \; 
e^{i \bm{\Delta}_T \cdot \bm{b}} \; f^a (x, b) ,
\\
-H^a (-x, \xi=0, t = -|\bm{\Delta}_T|^2) &= \int d^2 b \; 
e^{i \bm{\Delta}_T \cdot \bm{b}} \; \bar{f}^a (x, b) .
\end{align}
In this context the transverse charge density $\rho_1^p(b)$ represents
the integral over $x$ of the difference of the proton's quark and antiquark
distributions at transverse distance $b$, weighted by the quark charges $e_a$,
\begin{align}
& \rho_1^p(b) = \sum_a e_a \int_0^1 dx \; [f^a(x, b) - \bar f^a(x, b)],
\end{align}
which can be interpreted as the cumulative charge of the partons
in the proton at the transverse radius $b$. Equivalently, $\rho_1^p(b)$
represents the Fourier transform of the first moment of the charge-weighted GPDs,
\begin{align}
\int d^2 b \; e^{i \bm{\Delta}_T \cdot \bm{b}} \; \rho_1^p (b) 
&= \sum_a e_a \nonumber \int_{-1}^1 dx 
\nonumber
\\ & \times 
H^a (x, \xi=0, t = -|\bm{\Delta}_T|^2) .
\end{align}
A similar relation connects the density $\rho_2^p (b)$ with the proton helicity-flip GPDs
$E^a (x, \xi=0, t)$ \cite{Burkardt:2000za,Burkardt:2002hr}. The transverse densities are
thus directly related to the nucleon's transverse partonic structure in QCD.

The concepts of light-front quantization and partonic structure referenced here are used
only for the interpretation of the transverse densities but are not needed for their computation.
The densities are simple Fourier transforms of the invariant FFs and can be computed using hadronic
physics methods such as dispersion theory and effective field theory. It is this ``dual'' character
that makes the transverse densities so useful for nucleon structure studies. 
\subsection{Dispersive representation}
\label{subsec:dispersive_representation}
The nucleon FFs are analytic functions of $t$. The physical sheet has a principal cut
at positive real $t > t_{\rm thr}$; the threshold $t_{\rm thr}$ depends on the isospin channel
(see below). The FFs satisfy unsubtracted dispersion relations,
\begin{align}
F_i (t) \; &= \; \frac{1}{\pi}
\int_{t_{\rm thr}}^\infty dt' \, \frac{\textrm{Im}\, F_i(t')}{t' - t - i0} 
\hspace{2em} (i=1,2) ,
\label{dispersion_f12}
\end{align}
which express the functions at complex $t$ as integrals over their imaginary parts on the cut.
The real functions $\textrm{Im}\, F_i(t')$ are known as the spectral functions. They correspond
to processes in which the electromagnetic current with timelike momentum transfer $t > t_{\rm thr}$
couples to the nucleon through a hadronic state in the $t$-channel. These processes occur in
the unphysical region below the two-nucleon threshold, $t_{\rm thr} < t < 4 m_N^2$, where
the spectral functions cannot be measured directly and have to be constructed
using theoretical methods; see Ref.~\cite{Lin:2021umz} for a review.
In the isovector FFs the lowest-mass $t$-channel
state is two-pion state with $t_{\rm thr} = 4 M_\pi^2$; in the isoscalar FF it is the
three-pion state with $t_{\rm thr} = 9 M_\pi^2$.

The transverse densities Eqs.~(\ref{rho12_Fourier}) can be
computed as the Fourier transform of the dispersive representation of FFs,
Eq.~(\ref{dispersion_f12}) \cite{Strikman:2010pu,Miller:2011du}. One obtains a dispersive
(or spectral) representation of the densities as
\begin{align}
& \rho_1 (b) = \phantom{-} \frac{1}{2\pi^2} \int_{t_{\rm thr}}^{\infty} dt\ K_0(\sqrt{t}b)\
\textrm{Im}\, F_1(t) ,
\label{spectral_rho1}
\\
& \widetilde{\rho}_2 (b) = -\frac{1}{2\pi^2}
\int_{t_{\rm thr}}^{\infty} dt\ \frac{\sqrt{t}}{2 m_N}\ K_1(\sqrt{t}b)\ \textrm{Im}\, F_2(t) ,
\label{spectral_rho2t}
\end{align}
where $K_0$ and $K_1$ are the modified Bessel functions of the second kind.
Equations~(\ref{spectral_rho1}) and (\ref{spectral_rho2t}) express the densities at
a given distance $b$ as a superposition of contributions of $t$-channel states
(or exchanges) with squared mass $t$. The modified Bessel functions decay exponentially
at large arguments,
\begin{align}
K_{0, 1} (\sqrt{t} b) &\sim \sqrt{\frac{\pi}{2 \sqrt{t} b}}
\; e^{-\sqrt{t} b} 
\hspace{2em} (\sqrt{t} b \; \gg \; 1).
\label{K_n_large_argument}
\end{align}
The dispersive integrals for the densities therefore converge exponentially at large $t$,
and the integration effectively extends over masses
\begin{align}
\sqrt t &\sim 1/b .
\end{align}
This provides a mathematical formulation of the connection between the masses of the exchanges
and the ranges of the spatial distributions in the nucleon. In particular, the peripheral
densities at $b = \mathcal{O}(M_\pi^{-1})$ are governed by lowest-mass states in the dispersive
representation and can be computed and analyzed accordingly. Note that the actual spectral
composition of the densities --- how much the states with various $\sqrt{t}$ contribute to the
densities at given $b$ --- depends on the distribution of strength in the spectral functions
and can be established only with specific models of the latter.

The dispersive representation offers several theoretical and practical advantages for studying
the peripheral transverse densities compared to other approaches.
(a)~The dispersive representation permits efficient calculation of the peripheral densities.
The exponential convergence of the dispersion integrals reduces the contribution from the
high-mass region, where the spectral functions are poorly known. Calculations can focus on
the low-mass region --- the two-pion part of the isovector spectral function, for which
dedicated theoretical methods are available. The high-mass region can be parametrized
by effective poles, whose coefficients are fixed by sum rules; only the overall strength
in this region is relevant to the peripheral densities, not the details of the distribution.
(b) The dispersive representation automatically generates densities with the correct asymptotic
behavior at $b \rightarrow \infty$. The asymptotic behavior of the densities at $b \rightarrow \infty$
is governed by the analytic properties of the FFs in $t$ (position and strength of singularities),
which are explicitly realized in the dispersive representation. The densities exhibit an exponential
decay with a range governed by lowest-mass exchanges, modified by a pre-exponential factor
resulting from the behavior of the spectral function near the threshold 
\cite{Granados:2013moa}. The spectral integrals
Eqs.~(\ref{spectral_rho1}) and (\ref{spectral_rho2t}) permit stable numerical evaluation
of the densities in the region where they are exponentially small.\footnote{In
Ref.~\cite{Gramolin:2021gln} the low-$t$ nucleon form factors were analyzed by expanding
the transverse densities in a set of basis functions (orthogonal polynomials). That method produces
peripheral densities with an oscillating behavior, which is in conflict with the smooth exponential
fall-off dictated by the dispersive representation; see Ref.~\cite{Miller:2011du} and
the results of the present study, esp.\ Fig.~\ref{fig:rho12v_log}. The basis function expansion
does not naturally describe  the exponential smallness of the densities, and correlations between
many terms would be required to express it correctly.} In contrast, methods calculating
the densities as Fourier transforms of the FFs become numerically unstable at large $b$,
even if the FF parametrization have correct analytic properties.\footnote{FF parametrizations
with incorrect analytic properties, e.g. rational functions with singularities at complex $t$ with
$\textrm{Im}\, t \neq 0$ on the physical sheet, produce Fourier densities with qualitatively
incorrect asymptotic behavior at $b \rightarrow \infty$. Such FF parametrizations are principally
not adequate for evaluating densities at distances above $b \gtrsim 2$ fm, even if they provide good
fits to the spacelike form factor data at small $t < 0$; see Ref.~\cite{Miller:2011du}
for a discussion.} (c) The dispersive representation enables uncertainty estimates of the
peripheral densities. The densities generated by Eqs.~(\ref{spectral_rho1}) and (\ref{spectral_rho2t})
depend smoothly on the parameters of spectral function, even at large $b$ where they are
exponentially small. Varying the parameters of the low-mass spectral functions one can
estimate the uncertainties of the peripheral densities in a manner that respects analyticity
and is numerically stable. Methods using the frequency spectrum of the Fourier transform of FFs for
estimating the uncertainties of the densities are not appropriate for distances significantly
larger than 1 fm \cite{Venkat:2010by}.

The spectral functions obey certain integral relations (sum rules), which result from the constraints
on the nucleon FFs at $t = 0$ and their derivatives in the dispersive representation Eq.~(\ref{dispersion_f12}),
\begin{align}
\frac{1}{\pi}
\int_{t_{\rm thr}}^\infty dt \; \frac{\textrm{Im}\, F_1(t)}{t} &= Q ,
\label{sum_rule_f1_charge}
\\
\frac{1}{\pi}
\int_{t_{\rm thr}}^\infty dt \; \frac{\textrm{Im}\, F_1(t)}{t^2} &= \frac{1}{6} \langle r^2 \rangle_1
= \frac{1}{4} \langle b^2 \rangle_1 ,
\label{sum_rule_f1_radius}
\\
\frac{1}{\pi}
\int_{t_{\rm thr}}^\infty dt \; \frac{\textrm{Im}\, F_2(t)}{t}
&= \kappa , 
\label{sum_rule_f2_moment}
\\
\frac{1}{\pi} 
\int_{t_{\rm thr}}^\infty dt \; \frac{\textrm{Im}\, F_2(t)}{t^2}
&= \frac{1}{6} \kappa \langle r^2 \rangle_2
= \frac{1}{4} \kappa \langle b^2 \rangle_2 .
\label{sum_rule_f2_radius}
\end{align}
Additional relations follow from the asymptotic behavior of the nucleon FFs at large spacelike $|t|$,
\begin{align}
F_1(t) \sim |t|^{-2}, \hspace{2em} F_2(t) \sim |t|^{-3} \hspace{2em} (|t| \rightarrow \infty).
\label{F12_asymptotic}
\end{align}
This behavior is predicted by the QCD hard scattering mechanism up to logarithmic
corrections (counting rules) \cite{Lepage:1980fj}. In $F_1$ the predicted behavior is
approximately observed in data at $|t| \gtrsim$ 1 GeV$^2$; in $F_2$ the predicted
behavior is not observed at presently available momentum transfers \cite{JeffersonLabHallA:2001qqe};
see Ref.~\cite{Belitsky:2002kj} for a possible theoretical explanation in the context of QCD. 
When imposed on the dispersive representation of the FFs, Eq.~(\ref{F12_asymptotic}) implies
the relations (so-called superconvergence relations)
\begin{align}
\frac{1}{\pi}
\int_{t_{\rm thr}}^\infty dt \; \textrm{Im}\, F_1(t) &= 0,
\label{superconvergence_f1}
\\
\frac{1}{\pi}
\int_{t_{\rm thr}}^\infty dt \; \textrm{Im}\, F_2(t) &= 0,
\label{superconvergence_f2_0}
\\
\frac{1}{\pi}
\int_{t_{\rm thr}}^\infty dt \; t \; \textrm{Im}\, F_2(t) &= 0.
\label{superconvergence_f2_1}
\end{align}
At the level of the densities, these relations constrain the behavior in the limit $b \rightarrow 0$.
This can be derived from the dispersive representation, Eqs.~(\ref{spectral_rho1}) and
(\ref{spectral_rho2t}), by using the limiting behavior of the modified Bessel functions at small argument
(here $z \equiv \sqrt{t}b$),
\begin{align}
K_0(z) &= -\left( 1 + \frac{z^2}{4} + ...\right) \log z + \textrm{analytic},
\label{K0_small}
\\
K_1(z) &= \frac{1}{z} + \left(\frac{z}{2} + ... \right) \log z + \textrm{analytic},
\label{K1_small}
\end{align}
where ``analytic'' denotes the parts that are analytic at $z = 0$ (constants or positive powers).
Equation~(\ref{superconvergence_f1}) implies that
\begin{align}
\rho_1(b) \; \rightarrow \; \textrm{finite} \hspace{2em} \; \textrm{for} \;\; b \rightarrow 0.
\end{align}
The kernel $K_0(\sqrt{t}b)$ in Eq.~(\ref{spectral_rho1}) diverges logarithmically for
$b \rightarrow 0$ according to Eq.~(\ref{K0_small});
this would cause a logarithmic divergence of $\rho_1(b)$; however, the coefficient of the
divergent term is zero because of Eq.~(\ref{superconvergence_f1}).\footnote{Such
a logarithmic divergence at $b \rightarrow 0$ is observed in the charge density
in the pion, where the FF behaves as $F_\pi (t) \sim |t|^{-1}$ for $|t| \rightarrow \infty$
and no relation like Eq.~(\ref{superconvergence_f1})
exists \cite{Miller:2009qu,Miller:2010tz}. The finiteness of the
charge density in the nucleon at $b = 0$ appears natural in the parton picture,
as the nucleon is a more composite system than the pion.} Similarly, 
Eqs.~(\ref{superconvergence_f2_0}) and (\ref{superconvergence_f2_1})
imply that 
\begin{align}
\left.
\begin{array}{rcl}
\widetilde{\rho}_2(b) &\rightarrow& 0
\\[1ex]
\displaystyle
\frac{d\widetilde{\rho}_2}{db}(b) &\rightarrow& \textrm{finite}
\end{array}
\right\}
\;\; \textrm{for} \;\; b \rightarrow 0 .
\end{align}
\subsection{Spectral functions}
\label{subsec:method_spectral}
The spectral functions of the isovector nucleon FFs on the two-pion cut have been computed
using various theoretical approaches, such as analytic continuation of
pion-nucleon amplitudes \cite{Hohler:1976ax},
chiral EFT \cite{Gasser:1987rb,Bernard:1996cc,Kubis:2000zd,Kaiser:2003qp},
and Roy-Steiner equations \cite{Hoferichter:2016duk}. Here we employ the method of
DI$\chi$EFT, which combines general methods of dispersion theory (elastic unitarity
in the $\pi\pi$ channel, $N/D$ method) with specific dynamical input from chiral EFT.
The foundations of the method and its applications to FFs are described in detail in
Refs.~\cite{Alarcon:2017ivh,Alarcon:2017lhg,Alarcon:2018irp}. Here we summarize only the
main steps and the new features arising in the present application to densities.
The new features are: (a) We now construct the spectral functions $\textrm{Im}\, F_{1, 2}(t)$
as needed for the transverse densities, whereas in Refs.~\cite{Alarcon:2017lhg,Alarcon:2018irp}
we worked with $\textrm{Im}\, G_{E, M}(t)$. (b) We impose the superconvergence relations
resulting from the asymptotic behavior of $F_{1, 2}(t)$, which provide additional
constraints on the high-mass part of the spectral functions. (c) We implement a more flexible
parametrization of the high-mass part of the spectral function
to enable more realistic uncertainty estimates.

The isovector spectral functions are organized as
\begin{align}
\textrm{Im}\, F_i^V(t) \; &= \; 
\textrm{Im}\, F_i^V(t)[\pi\pi] \; \Theta (4 M_\pi^2 < t < t_{\rm max})
\nonumber \\[1ex]
 &+ \; 
\textrm{Im}\, F_i^V(t)[\textrm{high-mass}] \; \Theta (t > t_{\rm max})
\nonumber
\\[1ex]
&
(i = 1, 2),
\label{spectral_isovector_pipi_highmass}
\end{align}
where $\Theta$ is the step function (1 if the argument is true, otherwise 0).
The first term is the contribution of the two-pion cut that starts at $4 M_\pi^2$
and extends up to $t_{\rm max} \approx$ 1 GeV$^2$ (see below); this part is calculated theoretically.
The second term represents the contribution of high-mass states above $t_{\rm max}$
of unspecified hadronic composition; this part is parametrized through effective poles.

%
%
\begin{figure}
\includegraphics[width=.40\textwidth]{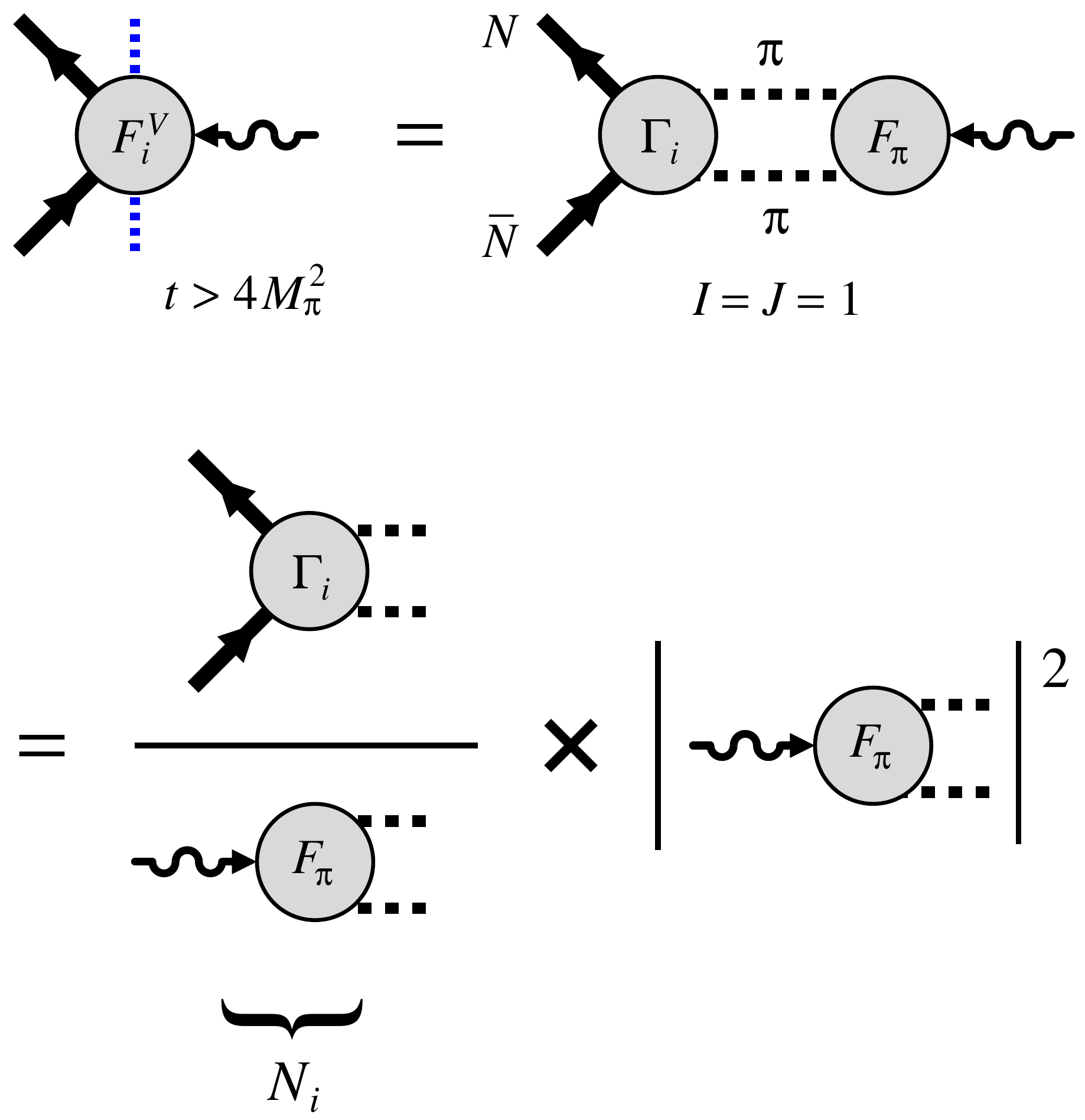}
\caption[]{Elastic unitarity relation for the two-pion part of the
isovector spectral functions $\textrm{Im}\, F_i^V(t)[\pi\pi]$ ($i = 1, 2$).
The first line shows the relation in the original form with the complex amplitudes,
Eqs.~(\ref{unitarity_complex}); the second line shows it in the explicitly real
form obtained with the $N/D$ method, Eq.~(\ref{unitarity_real}).}
\label{fig:unitarity}
\end{figure}
The two-pion spectral functions are obtained from the elastic unitarity relation
in the two-pion channel (see Fig.~\ref{fig:unitarity}). The relation is written
in a manifestly real form by applying the $N/D$
method \cite{Hohler:1974eq,Frazer:1960zza,Frazer:1960zzb}
\begin{align}
\textrm{Im}\, F_i^V(t)[\pi\pi]
&= \frac{k_{\rm cm}^3}{\sqrt{t}} \; \Gamma_i(t) \; F_\pi^*(t)
\label{unitarity_complex}
\\[1ex]
&= \frac{k_{\rm cm}^3}{\sqrt{t}} \; N_i (t) |F_\pi(t)|^2,
\label{unitarity_real}
\\[1ex]
k_{\rm cm} &\equiv \sqrt{t/4 - M_\pi^2},
\label{k_cm}
\\[1ex]
N_i (t) &\equiv \frac{\Gamma_i(t)}{F_\pi (t)} \hspace{2em} (i=1,2).
\label{N_def}
\end{align}
$k_{\rm cm}$ is the center-of-mass momentum of the $\pi\pi$ state in $t$-channel,
$\Gamma_i(t) (i = 1, 2)$ are the $\pi\pi \rightarrow N\bar{N}$
partial-wave amplitudes, and $F_\pi (t)$ is the pion timelike form factor.
The functions $\Gamma_i(t)$ and $F_\pi (t)$ in Eq.~(\ref{unitarity_complex})
are complex for $t > 4 M_\pi^2$ because of $\pi\pi$ rescattering but have the same phase.
The ratios $N_i(t) (i = 1, 2)$ in Eq.~(\ref{unitarity_real}) are real for $t > 4 M_\pi^2$
and possess only left-handed singularities; they are free of $\pi\pi$ rescattering effects
and describe the coupling of the $\pi\pi$ system to the nucleon. These function can be computed
in $\chi$EFT with good convergence. The $\pi\pi$ rescattering effects in Eq.~(\ref{unitarity_real})
are contained in the squared modulus of the pion form factor $|F_\pi(t)|^2$. This function is
measured in $e^+e^- \rightarrow \pi^+\pi^-$ exclusive annihilation experiments
and can be taken from a parametrization of the data \cite{Druzhinin:2011qd}.
In this way Eq.~(\ref{unitarity_real}) factorizes the rescattering effects in the $\pi\pi$
channel and permits chiral EFT-based computation of the spectral functions.
While Eq.~(\ref{unitarity_complex}) and Eq.~(\ref{unitarity_real}) are strictly valid
only up to the 4-pion threshold $t < 16\, M_\pi^2$, the $e^+e^-$ exclusive annihilation data 
indicate that 4-pion and other states are strongly suppressed up to $\sim$ 1 GeV$^2$
\cite{Druzhinin:2011qd}, so that the elastic unitarity relations can practically be used
up to $t_{\rm max} = 1$ GeV$^2$ \cite{Hohler:1974eq,Hohler:1976ax}.

We have computed the functions $N_i(t)$ in relativistic $\chi$EFT with $N$ and $\Delta$
intermediate states at LO and NLO accuracy. Contributions at N2LO accuracy have been estimated
assuming that the loop corrections have the same functional form as the tree-level result
(partial N2LO, or pN2LO approximation). The $\chi$EFT results for $N_i(t)$ can be obtained
from those of the functions $J^1_{\pm}(t)$, which appear in the $N/D$ representation of the
unitarity relation for the $G_{E, M}(t)$ form factors and were calculated in our
earlier studies \cite{Alarcon:2017lhg,Alarcon:2018irp}. The explicit formulas are
given in Appendix~\ref{app:spectral}.

The LO and NLO results for the functions $N_i(t)$ are given in terms of known low-energy constants
in the chiral Lagrangian. The pN2LO estimates involve one unknown parameter, $\lambda_i$, describing
the size of the loop corrections relative to the tree-level result. The total results for the
function $N_i(t)$ are thus given by $(i = 1,2)$
\begin{align}
& N_i(t) = N_i(t)[\textrm{LO}] + N_i(t)[\textrm{NLO}] + N_i(t)[\textrm{pN2LO}],
\label{spectral_isovector_pipi_orders}
\\[1ex]
& N_i(t)[\textrm{pN2LO}] \equiv \lambda_i \, N_i(t)[\textrm{N2LO-tree}].
\label{N_lambda}
\end{align}
The explicit form of $N_i(t)[\textrm{N2LO-tree}]$ is given in Eq.~(\ref{N_N2LO_tree}).
In summary, at pN2LO accuracy the two-pion part of each of the spectral functions in
Eq.~(\ref{spectral_isovector_pipi_highmass}) involves one free parameter, $\lambda_i$ ($i = 1, 2$).

The high-mass part of the isovector spectral functions in Eq.~(\ref{spectral_isovector_pipi_highmass})
is parametrized by effective poles. It is important to note that for our calculation of peripheral
densities we need only a summary description of the high-mass strength of the spectral function,
because of the strong numerical suppression of large $t$ in the dispersion integrals
Eqs.~(\ref{spectral_rho1}) and (\ref{spectral_rho2t}).
(see Sec.~\ref{subsec:dispersive_representation}).
We construct an appropriate parametrization by making reasonable assumptions about the positions
of the poles, treating variations of the position as part of the theoretical uncertainty,
and and fixing the strength of the poles through the dispersive sum rules
Eq.~(\ref{sum_rule_f1_charge}) et seq.\ and Eq.~(\ref{superconvergence_f1}) et seq.

Several observations suggest that the main strength of the isovector spectral functions
beyond the two-pion region is located around $t \approx$ 2\, GeV$^2$, and that higher
values of $t$ are strongly suppressed. (a)~The $e^+e^- \rightarrow$ hadrons exclusive annihilation
data show that the cross section at $t > 1$ GeV$^2$ is dominated by the $4\pi$ channel and 
concentrated around $t \approx 2$ GeV$^2$ \cite{Druzhinin:2011qd}. This suggests similar
behavior of the nucleon spectral functions, even if the connection with the annihilation
cross section is at the amplitude level and cannot be made explicit.
(b)~Dispersive fits to the spacelike nucleon FFs with flexible parametrizations
of the high-mass states using multiple effective poles find most of the strength
in the region around $t \approx$ 2\, GeV$^2$ \cite{Belushkin:2006qa,Lin:2021xrc}.
(c)~The dual resonance model describes the isovector spectral functions of the pion or nucleon FFs
through the exchange of vector resonances with masses $M^2_n = M_\rho^2 (1 + 2n)$,
with $M_0^2 \equiv M_\rho^2$. The first resonance after the $\rho$ has mass $M_1^2 = 3\, M_\rho^2 =$
1.8 GeV$^2$. If in the nucleon FFs the resonance contributions decrease rapidly with $n$,
the dominant contribution beyond the $\rho$ should come from this resonance.

Based on these observations, we parametrize the high-mass part of the isovector spectral
function $\textrm{Im}\, F_1^V(t)$ as
\begin{align}
&\textrm{Im}\, F_1^V(t)[\textrm{high-mass}]
\nonumber \\[1ex]
&= \pi a_1^{(V, 0)} \delta(t - t_1^{(V, 0)})
\nonumber \\[1ex]
&
+ \pi a_1^{(V, 1)} \delta'(t - t_1^{(V, 1)}) .
\label{spectral_isovector_highmass_f1}
\end{align}
The first term is a delta function, the second term is the derivative of a delta function.
The pole masses have the nominal value
\begin{align}
t_1^{(V, 0)}, t_1^{(V, 1)}[\textrm{nominal}] \; = \; 1.8 \, \textrm{GeV}^2 \; = \; 3 \, M_\rho^2 .
\label{pole_f1_nominal}
\end{align}
Their actual values are considered undetermined and will be allowed to vary randomly in a
plausible range; their distribution will be constrained by further physical requirements,
and the resulting variation in physical quantities will be regarded as a
theoretical uncertainty of the model (see Sec.~\ref{subsec:uncertainty}). Note that the
sum of the functions in Eq.~(\ref{spectral_isovector_highmass_f1}) parametrizes both
the ``strength'' and the ``shape'' of the high-mass spectral function in region around
$t \approx 2$ GeV$^2$ in an
effective form.\footnote{In dispersive fits of the nucleon isovector FFs, the high-mass states are
traditionally parametrized as a sum of simple delta functions at different positions. These parametrizations
give clusters of poles around $\sim 2$ GeV$^2$ with varying signs and unnaturally large
coefficients $\gg 1$ \cite{Belushkin:2006qa}, which can effectively be combined to a sum of
a single delta function and delta function derivatives. In this sense our parametrization
Eq.~(\ref{spectral_isovector_highmass_f1}) is equivalent to the traditional sum of
simple delta functions. An advantage of our parametrization is that all coefficients have
natural size $\sim 1$, and that one can directly implement the feature that all the ``structures''
arise from the same mass region.}

Combining the two-pion part given by Eqs.~(\ref{unitarity_real}), (\ref{spectral_isovector_pipi_orders}),
and (\ref{N_lambda}), and the high-mass part given by Eq.~(\ref{spectral_isovector_highmass_f1}),
the isovector spectral function $\textrm{Im}\, F_1^V(t)$ contains three unknown parameters:
\begin{align}
\lambda_1, \; a_1^{(V, 0)}, \; a_1^{(V, 1)}.
\label{parameters_f1}
\end{align}
We fix these parameters through the sum rules for the isovector charge and radius,
see Eqs.~(\ref{sum_rule_f1_charge}) and (\ref{sum_rule_f1_radius}),
and the superconvergence relation for $F_1^V$, see Eq.~(\ref{superconvergence_f1}):
\begin{align}
\frac{1}{\pi}
\int_{t_{\rm thr}}^\infty dt \; \frac{\textrm{Im}\, F_1^V(t)}{t} &= Q^V,
\label{sumrules_isovector_f1}
\\
\frac{1}{\pi}
\int_{t_{\rm thr}}^\infty dt \; \frac{\textrm{Im}\, F_1^V(t)}{t^2} &= \frac{1}{6} \langle r^2 \rangle_1^V,
\\
\frac{1}{\pi}
\int_{t_{\rm thr}}^\infty dt \; \textrm{Im}\, F_1^V(t) &= 0,
\label{sumrules_isovector_f1_end}
\end{align}
where
\begin{align}
Q^V &\equiv \frac{1}{2}(Q^p - Q^n) \; = \; \frac{1}{2},
\label{sumrules_isovector_f1_values}
\\[1ex]
\langle r^2 \rangle_1^V
&\equiv \frac{1}{2}
\left( \langle r^2 \rangle_1^p - \langle r^2 \rangle_1^n \right).
\label{sumrules_isovector_f1_values_end}
\end{align}
The relations Eqs.~(\ref{sumrules_isovector_f1})--(\ref{sumrules_isovector_f1_end})
constrain weighted integrals of the total isovector
spectral function $\textrm{Im}\, F_1^V(t)$. The integrals extend over the two-pion and the high-mass
parts of the spectral functions (here $n = 0, 1, 2$),
\begin{align}
& \int_{t_{\rm thr}}^\infty dt \; t^{-n} \;
\textrm{Im}\, F_1^V(t)
\nonumber \\[1ex]
\equiv &
\int_{t_{\rm thr}}^{t_{\rm max}} dt \; t^{-n} \;
\textrm{Im}\, F_1^V(t)[\pi\pi]
\nonumber \\[1ex]
+& \int_{t_{\rm max}}^\infty dt \; t^{-n} \;
\textrm{Im}\, F_1^V(t)[\textrm{high-mass}].
\end{align}
The integral over the two-pion part is a continuous integral and computed numerically;
the integral over the high-mass part is a sum of delta function derivative integrals and
computed exactly. Since the integrands depend linearly on the parameters Eq.~(\ref{parameters_f1}),
one obtains a system of linear equations for the parameters, which can easily be solved.

In the spectral function $\textrm{Im}\, F_2^V(t)$, we parametrize the high-mass part as
\begin{align}
&\textrm{Im}\, F_2^V(t)[\textrm{high-mass}]
\nonumber \\[1ex]
&= \pi a_2^{(V, 0)} \delta(t - t_2^{(V, 0)})
\nonumber \\[1ex]
&+ \pi a_2^{(V, 1)} \delta'(t - t_2^{(V, 1)})
\nonumber \\[1ex]
&+ \pi a_2^{(V, 2)} \delta''(t - t_2^{(V, 2)}) ,
\label{spectral_isovector_highmass_f2}
\end{align}
which compared to Eq.~(\ref{spectral_isovector_highmass_f1}) includes also a term
with a second derivative of a delta function. The three pole masses again have the nominal value
\begin{align}
t_2^{(V, 0)}, t_2^{(V, 1)}, t_2^{(V, 2)}[\textrm{nominal}] \; = \; 1.8 \, \textrm{GeV}^2 \; = \; 3 \, M_\rho^2 ,
\label{pole_f2_nominal}
\end{align}
and will be allowed to vary randomly in an interval around this value (see Sec.~\ref{subsec:uncertainty}).
The combined spectral function now contains four unknown parameters:
\begin{align}
\lambda_2, \; a_2^{(V, 0)}, \; a_2^{(V, 1)}, \; a_2^{(V, 2)} .
\label{parameters_f2}
\end{align}
They are fixed through the sum rules for the isovector magnetic moment and the magnetic radius,
see Eqs.~(\ref{sum_rule_f2_moment}) and (\ref{sum_rule_f2_radius}), and by the two superconvergence
relations for $F_2$, see Eqs.~(\ref{superconvergence_f2_0}) and (\ref{superconvergence_f2_1}):
\begin{align}
\frac{1}{\pi}
\int_{t_{\rm thr}}^\infty dt \; \frac{\textrm{Im}\, F_2^V(t)}{t}
&= \kappa^V ,
\label{sumrules_isovector_f2}
\\
\frac{1}{\pi} 
\int_{t_{\rm thr}}^\infty dt \; \frac{\textrm{Im}\, F_2^V(t)}{t^2}
&= \frac{1}{6}\langle r^2 \rangle_2^V ,
\\
\frac{1}{\pi}
\int_{t_{\rm thr}}^\infty dt \; \textrm{Im}\, F_2^V(t) &= 0 ,
\\
\frac{1}{\pi}
\int_{t_{\rm thr}}^\infty dt \; t \; \textrm{Im}\, F_2^V(t) &= 0 ,
\label{sumrules_isovector_f2_end}
\end{align}
where
\begin{align}
\kappa^V &\equiv \frac{1}{2}(\kappa^p - \kappa^n) ,
\label{sumrules_isovector_f2_values}
\\[1ex]
\langle r^2 \rangle_2^V
&\equiv 
\frac{1}{2}\left( \kappa^p \langle r^2 \rangle_2^p - \kappa^n \langle r^2 \rangle_2^n \right) .
\label{sumrules_isovector_f2_values_end}
\end{align}

In our approach the unknown parameters in the spectral functions are fixed by the dispersive
sum rules and expressed in terms of the values and derivatives of the FFs at $t = 0$.
Since values of the FFs (charges and magnetic moments) are known, this leaves the derivatives
(radii) as the effective parameters of our model. With the spectral functions determined by
the radii, our approach can then predict the spacelike FFs and the densities in terms
of the radii. This particular ``information flow'' is made possible by the analytic properties
of the FFs, which relate integrals over the spectral functions to derivatives of the FF at $t = 0$.

%
%
\begin{figure}[t]
\includegraphics[width=.45\textwidth]{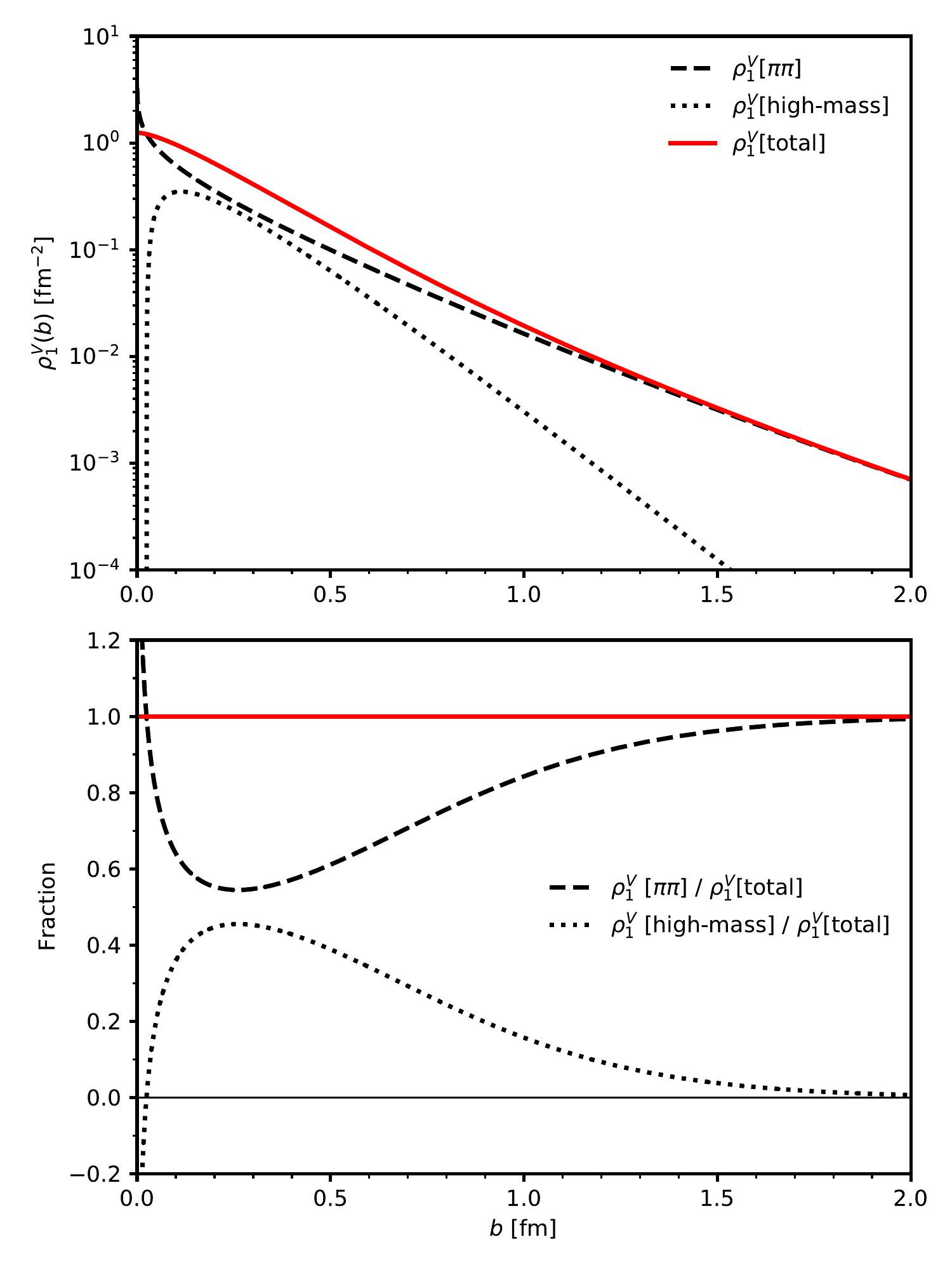}
\caption[]{Contributions of the two-pion and high-mass parts of the spectral function $\textrm{Im}\, F_1^V(t)$,
Eq.~(\ref{spectral_isovector_pipi_highmass}), to the dispersion integral of the density $\rho_1^V(b)$
(nominal parameters). {\it Upper panel:} Absolute contributions. Dashed line: $\rho_1^V(b)[\pi\pi]$.
Dotted line: $\rho_1^V(b)[\textrm{high-mass}]$. Solid line: $\rho_1^V(b)[\textrm{total}]$.
{\it Lower panel:} Relative contributions. Same as in upper panel, but divided by
total density $\rho_1^V(b)[\textrm{total}]$.}
\label{fig:rho1v_frac}
\end{figure}
When computing the densities in the dispersive representation, Eqs.~(\ref{spectral_rho1}) and
(\ref{spectral_rho2t}), the integrals receive contributions from the two-pion and the high-mass
parts of the spectral functions, Eq.~(\ref{spectral_isovector_pipi_highmass}). An important question
is how the relative contributions depend on the distance $b$, which represents an external parameter
in the integrals. Figure~\ref{fig:rho1v_frac} shows the two-pion and high-mass contributions
to $\rho_1^V(b)$ obtained with our spectral functions (here, for the the nominal parameter values).
The top panel shows the absolute contributions; the bottom panel shows the relative contributions,
i.e., the fractions of $\rho_1^V(b)$ due to the two-pion and high-mass parts. One observes that
in $\rho_1^V(b)$ the two-pion part accounts for $>80\%$ of at $b > 1$ fm, and
$>60\%$ at $b > 0.5$ fm. Similar relative contributions are found in the dispersion integral for
$\widetilde\rho_2^V(b)$ (not shown in the figure). In $\widetilde\rho_2^V(b)$ the two-pion part
accounts for $>97\%$ of at $b > 1$ fm, and $>50\%$ at $b > 0.5$ fm. 
These findings are central to our approach, as they quantify the dominance
of the two-pion state at large distances and justify the summary description of the high-mass states
for the purpose of computing the peripheral densities.

In the present study our focus is on the isovector channel, where the two-pion state in the spectral
functions generates the dominant contributions to the peripheral nucleon densities. In order to
compute the individual proton and neutron densities in the dispersive representation, we need
also the isoscalar spectral function. A parametrization of the isoscalar spectral functions,
constructed along similar lines as for the isovector spectral function but relying more on
empirical information, is described in Appendix~\ref{app:isoscalar}.
\subsection{Uncertainty estimates}
\label{subsec:uncertainty}
Our dispersive approach allows us to estimate the uncertainties of the spectral functions
and the densities obtained from them. We consider two sources of uncertainties:

{\it (I) Uncertainties due to the parametrization of the high-mass part of the spectral functions.}
The high-mass part of the isovector spectral function is linked to the low-mass part through
the dispersive sum rules, Eqs.~(\ref{sumrules_isovector_f1})--(\ref{sumrules_isovector_f1_end})
and Eqs.~(\ref{sumrules_isovector_f2})--(\ref{sumrules_isovector_f2_end}).
The parametrization of the high-mass part can therefore influence the low-mass part of spectral
functions and indirectly affect observables sensitive to the low-mass part, such as the peripheral densities.
We estimate this uncertainty by varying the positions of the high-mass poles in the
isovector spectral functions, Eqs.~(\ref{pole_f1_nominal}) and Eqs.~(\ref{pole_f2_nominal}).
As the plausible range of variation we consider
\begin{align}
\left.
\begin{array}{r}
t_1^{(V, 0)}, t_1^{(V, 1)} \\[1ex]
t_2^{(V, 0)}, t_2^{(V, 1)}, t_2^{(V, 2)}
\end{array}
\right\} &= (1.2 - 2.4) \; \textrm{GeV}^2 \, = \, (2 - 4) \, M_\rho^2 .
\label{variation_highmass}
\end{align}
This range allows for variations of the pole masses with a maximum/minimum ratio of 2, which is a very
significant change. Eq.~(\ref{variation_highmass}) covers the entire region of the secondary
peak of the $e^+e^-$ annihilation cross section above the $\rho$ resonance \cite{Druzhinin:2011qd}.
In the context of the dual resonance model, Eq.~(\ref{variation_highmass}) corresponds to varying
the pole position from the $n = 1$ resonance at $3 M_\rho^2$ to values that are half way between
this one and the $n = 0$ or 2 resonances. Note that we let the mass parameters in the delta functions
vary independently of each other over the given range, so that the parametrization represents
a wide range of ``shapes'' of the spectral function.

We further constrain the set of mass parameters by requiring that the variation of the spacelike
form factor generated by the spectral function be within a certain range around the nominal value.
This is essentially a stability condition, which eliminates extreme values of the mass parameters
that would lead to large excursions of the spacelike form factor and can be ruled out on physical grounds.
We implement this by requiring that (here $i=1, 2$)
\begin{align}
\frac{F_i^V (t_{\rm ref})[\textrm{varying pole masses}]}
{F_i^V (t_{\rm ref})[\textrm{nominal}]} \; - \; 1 \; < \; \epsilon ,
\label{stability}
\end{align}
where $t_{\rm ref} < 0$ is a spacelike $t$ value. In the following applications we choose
$t_{\rm ref} = -1$ GeV$^2$ and $\epsilon = 0.1$ for both $F_1^V$ and $F_2^V$; the choice is
justified in the following; other choices are possible. The parameter variation Eq.~(\ref{variation_highmass}),
supplemented by the stability condition Eq.~(\ref{stability}), generates a functional variation in the
high-mass part of the spectral function which we regard as the theoretical uncertainty of our model.
Note that the stability condition Eq.~(\ref{stability}) restricts the variation of the theoretical FF
prediction relative to the nominal value of the model, not relative to an experimental value;
no fitting to the FF data is performed here. The parameters $t_{\rm ref}$ and $\epsilon$ are chosen such
that the resulting theoretical model uncertainty of the FFs is reasonable and covers the experimental data.
In this way the experimental FF data are used only in estimating the theoretical uncertainty
of the model, not in determining the nominal model prediction.

%
%
\begin{figure}[t]
\includegraphics[width=.49\textwidth]{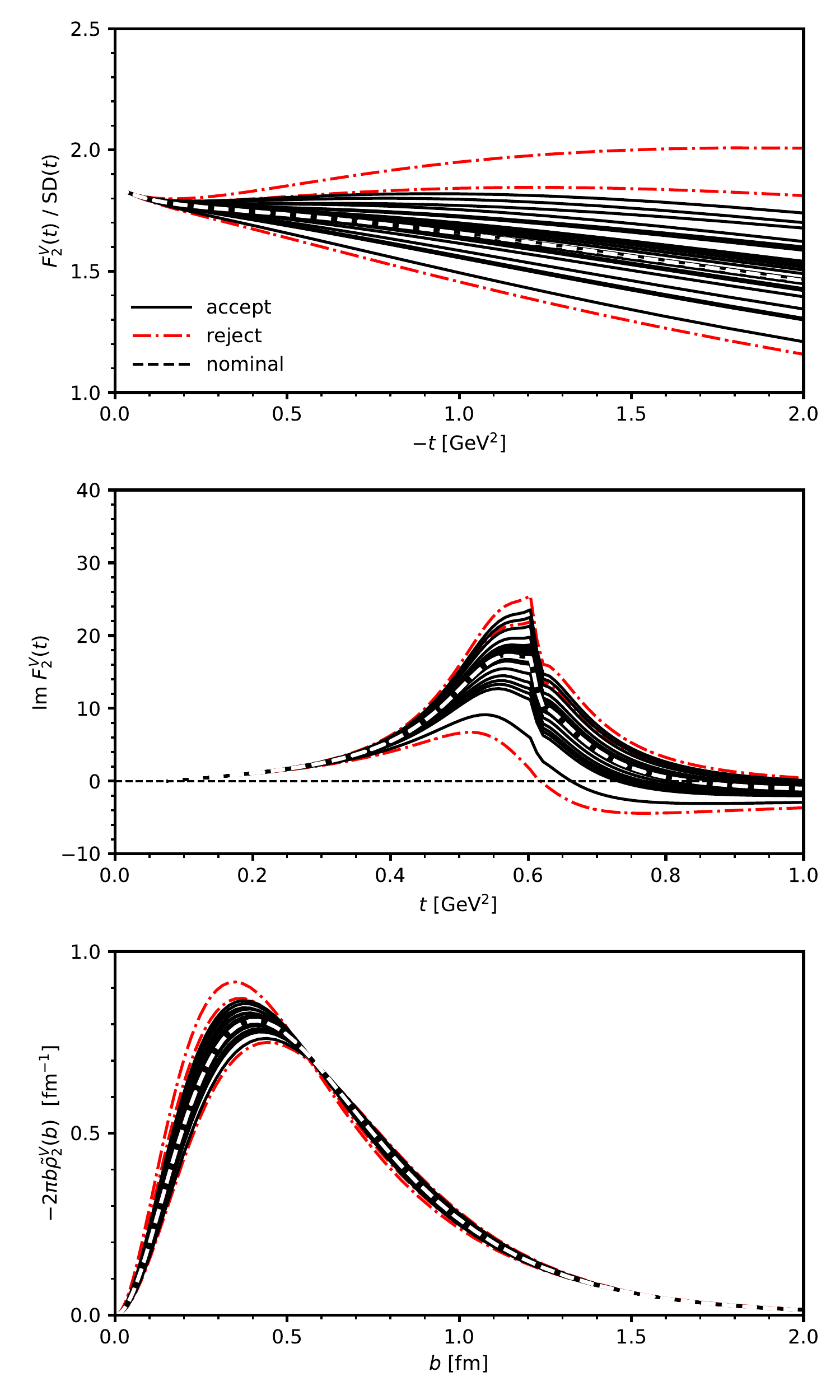}
\caption[]{Illustration of the procedure for estimating the uncertainty resulting from the high-mass
part of the spectral function $\textrm{Im}\, F_2^V(t)$. A random ensemble of parameters for the
high-mass pole positions Eq.~(\ref{variation_highmass}) is generated. With these parameters, the
spectral function and the spacelike FF are computed. The parameter ensemble is then restricted
to values for which the spacelike FF satisfies the stability condition Eq.~(\ref{stability})
(see upper panel, accept/reject). The restricted parameter ensemble is then used to evaluate the
uncertainty of the spectral function (middle panel) and the density (lower panel). In all panels, the functions
with accepted parameters by Eq.~(\ref{stability}) are shown as solid black lines, the ones with
rejected parameters as dot-dashed red lines. The functions with the nominal parameters are
shown by the dashed white line.}
\label{fig:rand_comb}
\end{figure}
To map out the theoretical uncertainty in practice, we generate a random ensemble of mass parameters
in the range of Eq.~(\ref{variation_highmass}) and retain those for which the spacelike FFs
satisfy the condition Eq.~(\ref{stability}). We then use this restricted ensemble to generate
uncertainty bands in the spectral functions and transverse densities (and possibly other quantities
derived from the spectral functions). Figure~\ref{fig:rand_comb} illustrates the procedure
in the case of $F_2^V$. One observes that the procedure generates natural uncertainty bands,
which are approximately symmetric around the nominal value. The resulting uncertainty will be quoted
as ``high-mass uncertainty'' in the results below.

We emphasize that the procedure respects analyticity and the dispersive sum rules.
Each instance in the ensemble corresponds to a form factor with correct analyticity in $t$,
and a density with correct asymptotic behavior at large $b$. Each instance represents
a spectral function that satisfies the sum rules
Eqs.~(\ref{sumrules_isovector_f1})--(\ref{sumrules_isovector_f1_end}) and
Eqs.~(\ref{sumrules_isovector_f2})--(\ref{sumrules_isovector_f2_end})
and produces form factors and densities with the correct normalization.
The only differences between the instances are in the form of the high-mass spectral function,
and in the distribution of strength between the low-mass and high-mass regions.

{\it (II) Uncertainties due to the nucleon radii.} The nucleon radii determine the spectral
function parameters through the dispersive sum rules
Eqs.~(\ref{sumrules_isovector_f1})--(\ref{sumrules_isovector_f1_end}) and
Eqs.~(\ref{sumrules_isovector_f2})--(\ref{sumrules_isovector_f2_end}).
We can estimate the resulting uncertainty by varying the
value of the radii. The empirical values of the radii and their uncertainties
are summarized in Appendix~\ref{app:radii}. For our uncertainty estimate, we vary the the
radii in a range corresponding to their empirical uncertainty
\begin{align}
\langle r^2 \rangle_1^V &\rightarrow \langle r^2 \rangle_1^V [\textrm{nominal}]
\times (1 \pm 0.03),
\label{variation_radius_1}
\\
\langle r^2 \rangle_2^V &\rightarrow \langle r^2 \rangle_2^V [\textrm{nominal}]
\times (1 \pm 0.02).
\label{variation_radius_2}
\end{align}
We then follow the effect of this variation from the spectral function to the form factors
and densities calculated as dispersive integrals. The resulting uncertainty will be quoted
as ``radius uncertainty'' in the results below.
\section{Results}
\label{sec:results}
\subsection{Spectral functions}
\label{subsec:results_spectral}
%
%
\begin{figure*}[t]
\begin{tabular}{ll}
\includegraphics[width=.45\textwidth]{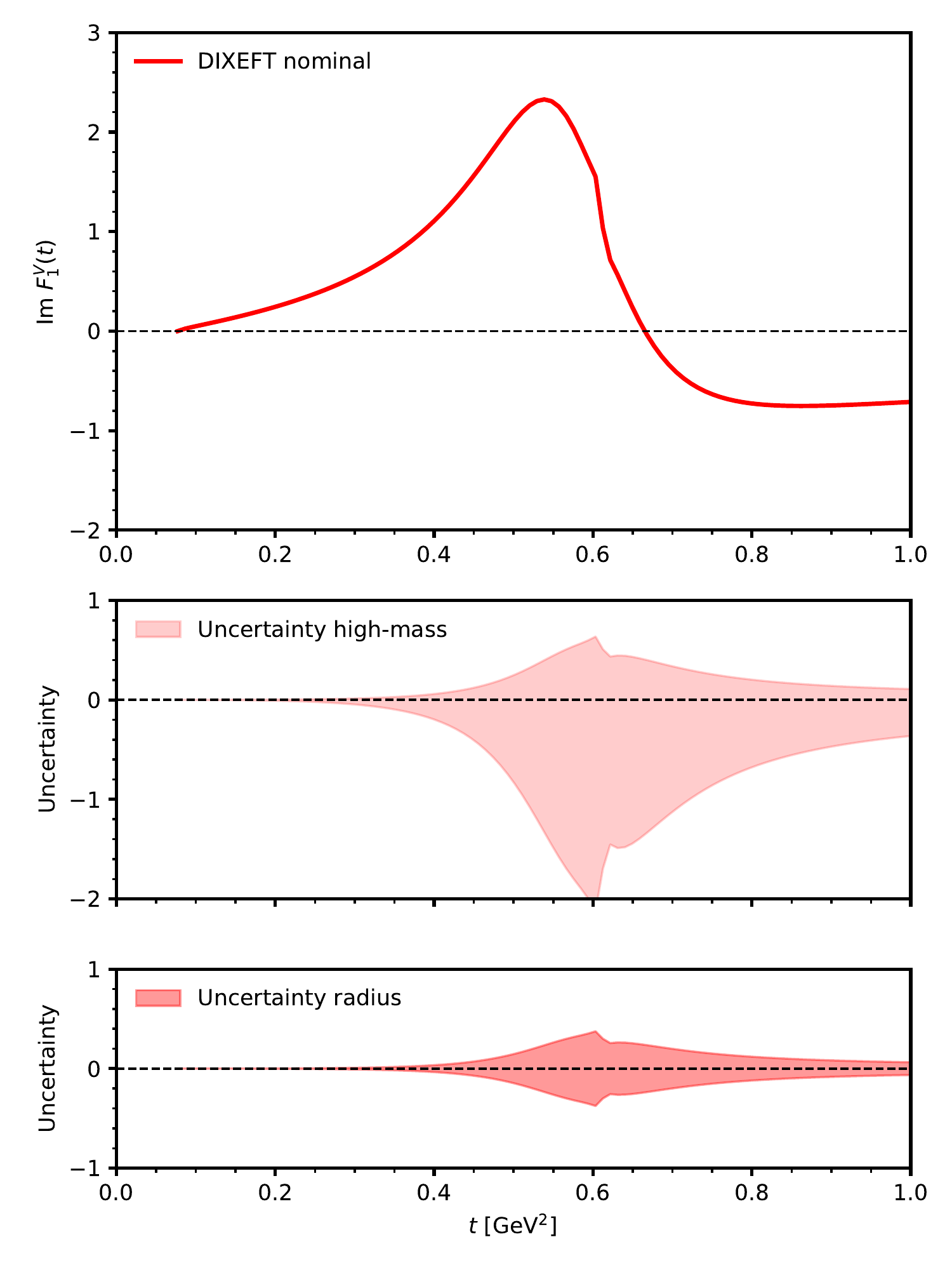} &
\includegraphics[width=.45\textwidth]{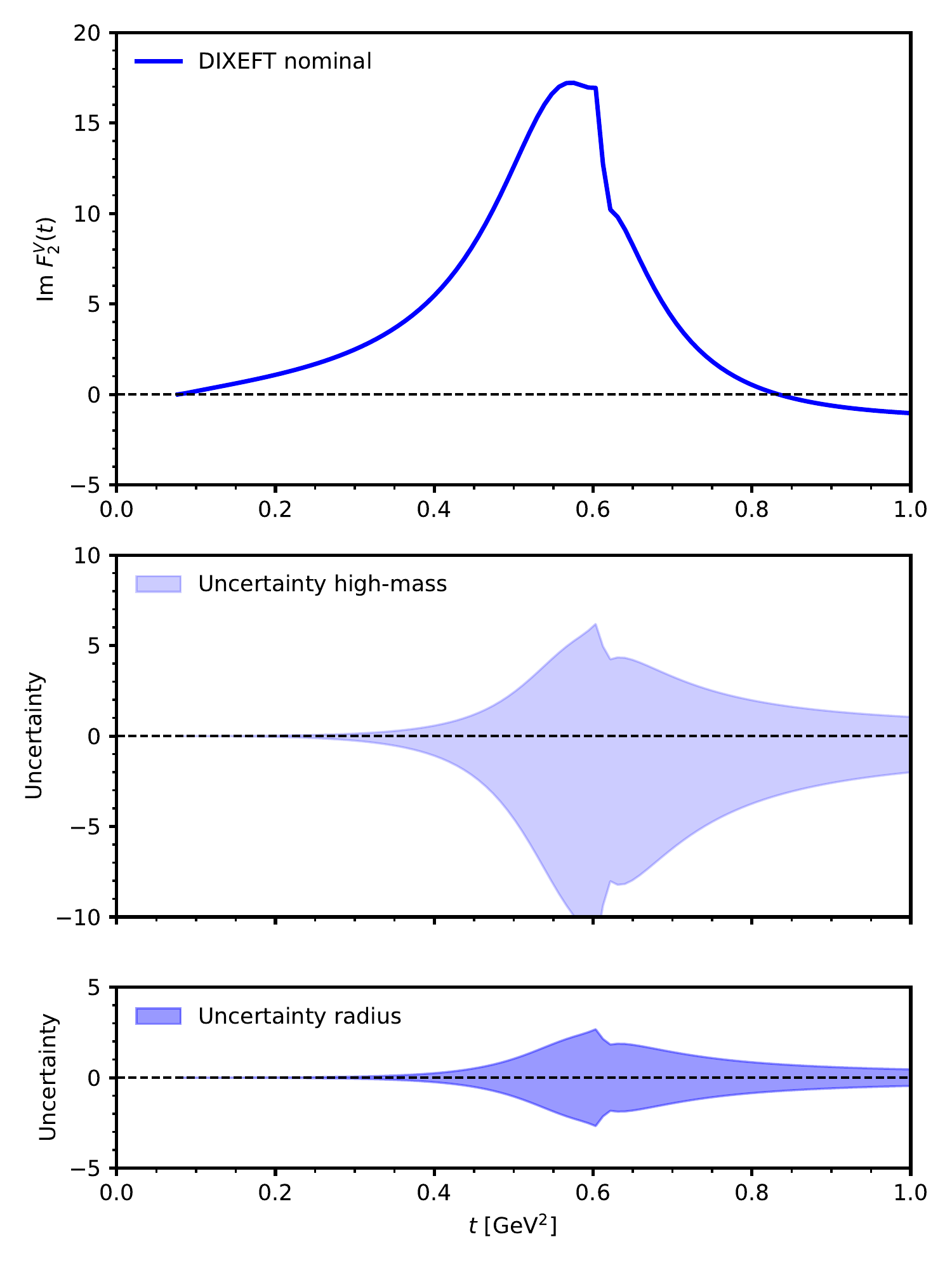}
\\[-3ex]
\end{tabular}
\caption[]{Two-pion part of isovector spectral functions $\textrm{Im}\, F_1^V(t)$ (left column)
and $\textrm{Im}\, F_2^V(t)$ (right column) obtained in DI$\chi$EFT. In each column:
{\it Upper panel:} DI$\chi$EFT result with nominal parameters.
{\it Middle panel:} Uncertainties resulting from the parametrization of high-mass states.
{\it Lower panel:} Uncertainties resulting from the nucleon isovector radii.}
\label{fig:spectral}
\end{figure*}
%
%
\begin{figure*}[t]
\begin{tabular}{ll}
\includegraphics[width=.45\textwidth]{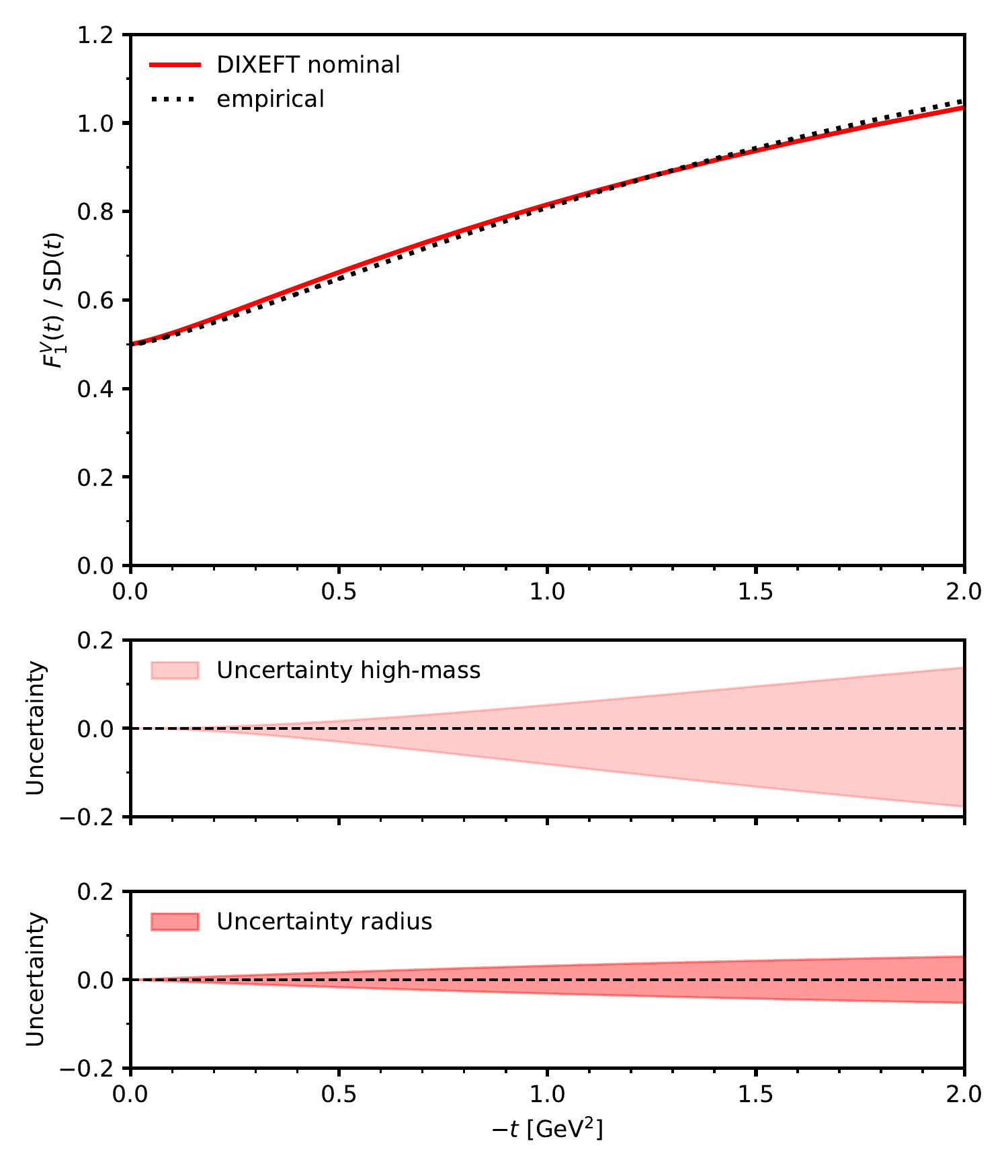} &
\includegraphics[width=.45\textwidth]{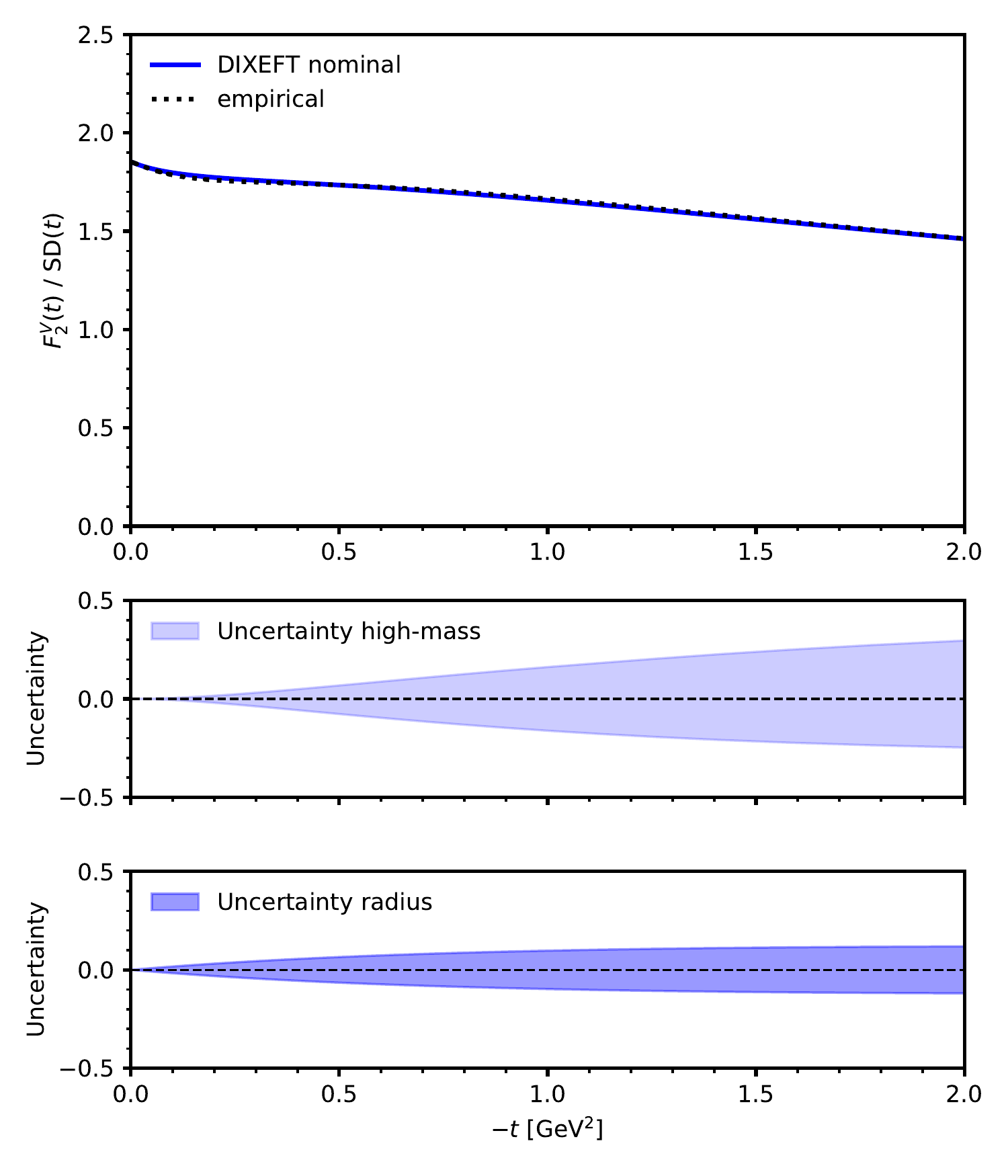}
\\[-3ex]
\end{tabular}
\caption[]{Isovector nucleon form factors $F_1^V(t)$ (left column) and $F_2^V(t)$ (right column)
at spacelike $t < 0$ obtained from the DI$\chi$EFT spectral functions.
In each column: {\it Upper panel, solid line:} DI$\chi$EFT result with nominal parameters.
{\it Middle panel:} Uncertainty resulting from the parametrization of high-mass states.
{\it Lower panel:} Uncertainty resulting from the nucleon isovector radii.
{\it Upper panel, dotted line:} Empirical FFs \cite{Ye:2017gyb}.
All FFs are shown divided by the standard dipole FF.}
\label{fig:f1v_f2v}
\end{figure*}
Figure~\ref{fig:spectral} shows our results for the two-pion part of the isovector spectral
functions $\textrm{Im}\, F_1^V(t)$ and $\textrm{Im}\, F_2^V(t)$ and their uncertainties
obtained with the methods of Sec.~\ref{subsec:method_spectral} and \ref{subsec:uncertainty}.
The upper panels show the spectral functions obtained
with the nominal parameters for the high-mass poles and the radii. One observes:
(a) The spectral functions show the characteristic peak from the $\rho$ resonance in
the $\pi\pi$ channel. This essential feature arises through the pion timelike form factor
in the elastic unitarity relation Eq.~(\ref{unitarity_real}). (b) Both spectral functions
(with the nominal parameters) change sign and become negative above the $\rho$ region.

The middle and lower panels of Fig.~\ref{fig:spectral} show the uncertainties in the two-pion part
of the spectral functions resulting from the parametrization of the high-mass part and from the
nucleon radii, estimated with the procedure of Sec.~\ref{subsec:uncertainty}.
(Note that the figure shows only the variation of the two-pion part of the spectral function;
the high-mass part undergoes a corresponding variation with the parameters, so that the
sum rule are satisfied; this part is not shown in the figure.) One observes:
(a) The uncertainties of the spectral functions are negligible in the region of $t$ from
the threshold at $4 M_\pi^2$ to $\sim 0.4$ GeV$^2$. In this region the chiral expansion of
the functions $N_i(t)$ is well convergent, and the spectral functions represent genuine
predictions of the theory. Notice that the enhancement of the spectral functions
through $\pi\pi$ rescattering is already very significant in this region \cite{Alarcon:2017asr}.
(b) In the region of the $\rho$ resonance, the spectral functions show significant uncertainties
from the high-mass states and from the radii. The behavior is this region is mainly
constrained by the sum rules, so that the spectral functions  become sensitive to the parameters,
as expected. The relative uncertainties are of order unity and approximately the same
in $\textrm{Im}\, F_1^V(t)$ and $\textrm{Im}\, F_2^V(t)$.

In the present study we use the spectral functions to compute the peripheral densities.
The dispersive integrals for the peripheral densities converge rapidly and sample mostly the
two-pion part of the spectral functions; the contribution of high-mass states is strongly suppressed
(see Fig.~\ref{fig:rho1v_frac}).
Our DI$\chi$EFT method and uncertainty estimates aim to provide a realistic description of the
two-pion part, while parametrizing the high-mass part in summary form. The dispersive integrals for
the spacelike FFs converge more slowly and are more sensitive to the high-mass states.
The computation of spacelike FFs therefore generally places stronger demands on the description
of the high-mass states than are needed in the present study. Still, it is instructive to see how
our simple spectral functions perform in the computation of the spacelike form factors,
for which experimental data are available.

Figure~\ref{fig:f1v_f2v} shows the spacelike FFs $F_1(t)$ and $F_2(t)$ obtained with our spectral functions.
[The plots show the FFs divided by the standard dipole FF
$\textrm{SD}(t) \equiv (1 - t/0.71 \, \textrm{GeV}^2)^{-2}$.]
The top panels show the predictions with the nominal values of the high-mass pole position
and the nucleon radii. One observes that the nominal predictions agree very
well with the empirical FFs extracted from experimental data \cite{Ye:2017gyb}. This indicates
that our assumptions made in parametrizing the high-mass part of the spectral function
(in particular, the rapid saturation at low masses $t \approx 3 M_\rho^2$) are realistic
at the quantitative level. The middle and lower panels show the uncertainties of the
predictions due to the position of the high-mass poles and the values of the
nucleon radii, estimated with the procedure of Sec.~\ref{subsec:uncertainty}.
One observes that the procedure gives a reasonable uncertainty estimate of the spacelike FFs
that is approximately symmetric around the nominal value (by design of the procedure)
and covers the experimental values. We emphasize that our goal here is not to predict or
analyze the spacelike FFs, but just to validate that our spectral function results are
compatible with the spacelike FF data. It is clear that a much more accurate description of the
spacelike FFs could be achieved within our framework if the value of the high-mass poles
were used as fit parameters, as is commonly done in dispersive fits \cite{Belushkin:2006qa,Lin:2021xrc}.
\subsection{Isovector densities}
\label{subsec:isovector_densities}
%
%
\begin{figure*}[!]
\begin{tabular}{ll}
\includegraphics[width=.43\textwidth]{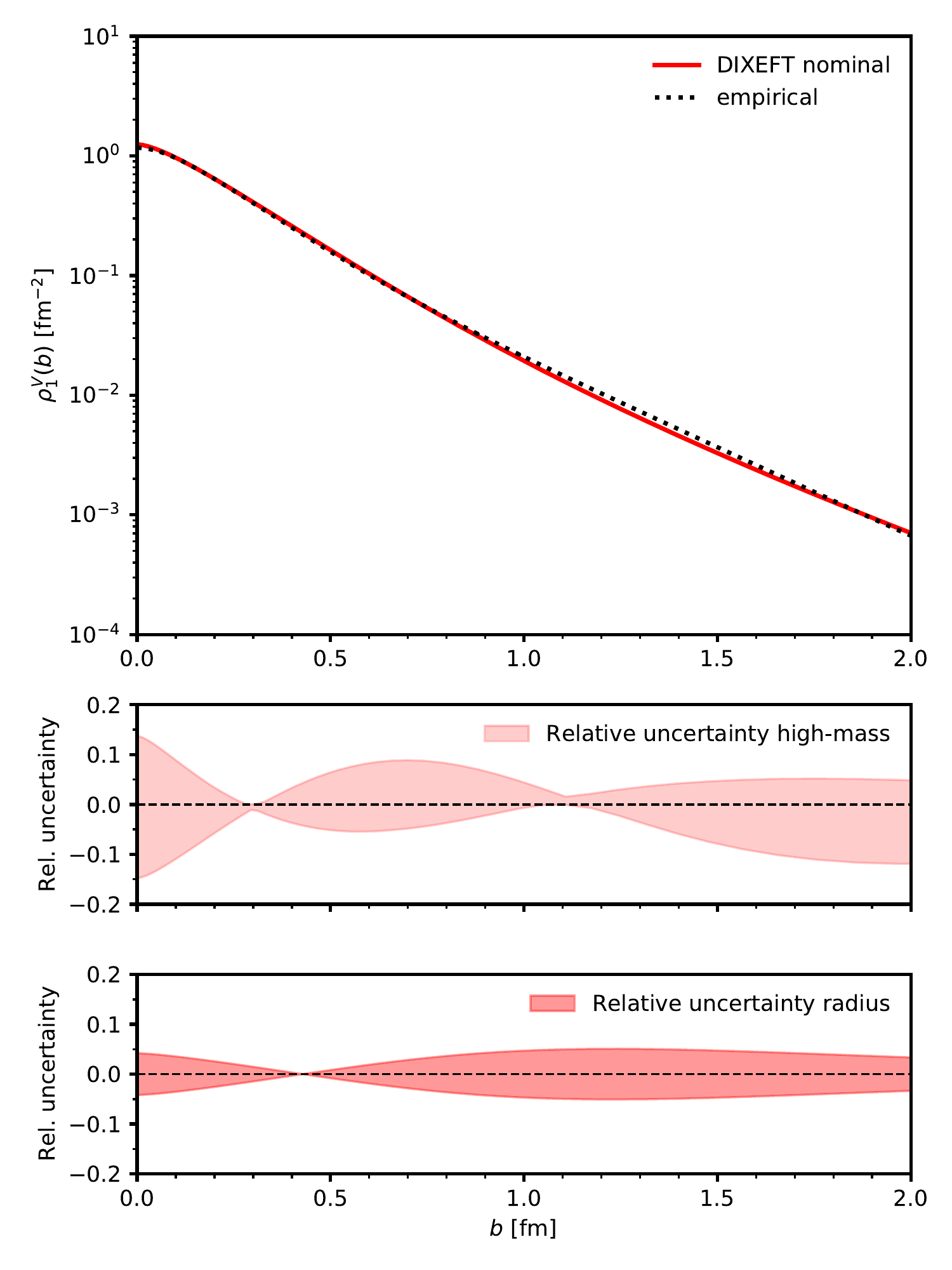} &
\includegraphics[width=.43\textwidth]{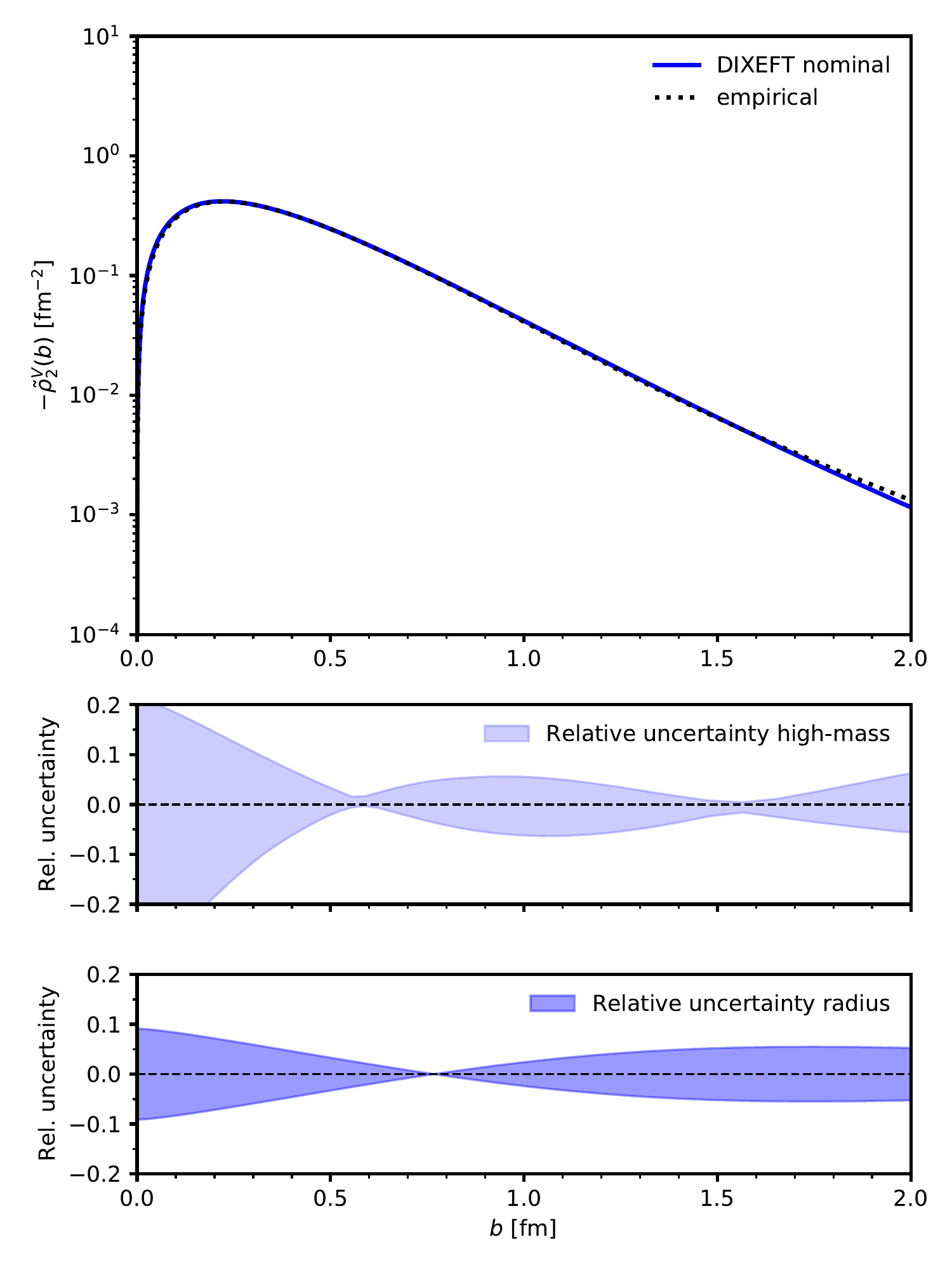}
\\[-3ex]
\end{tabular}
\caption[]{Isovector densities $\rho_1^V(b)$ (left column) and $-\widetilde{\rho}_2^V(b)$ (right column)
and their relative uncertainties obtained with the DI$\chi$EFT spectral functions. In each column:
{\it Upper panel, solid line:} DI$\chi$EFT prediction with nominal parameters.
{\it Middle panel:} Relative uncertainty from high-mass states.
{\it Lower panel:} Relative uncertainty from nucleon isovector radii.
{\it Upper panel, dotted line:} Density from empirical FFs \cite{Ye:2017gyb}.}
\label{fig:rho12v_log}
\end{figure*}
%
%
\begin{figure*}[!]
\begin{tabular}{ll}
\includegraphics[width=.43\textwidth]{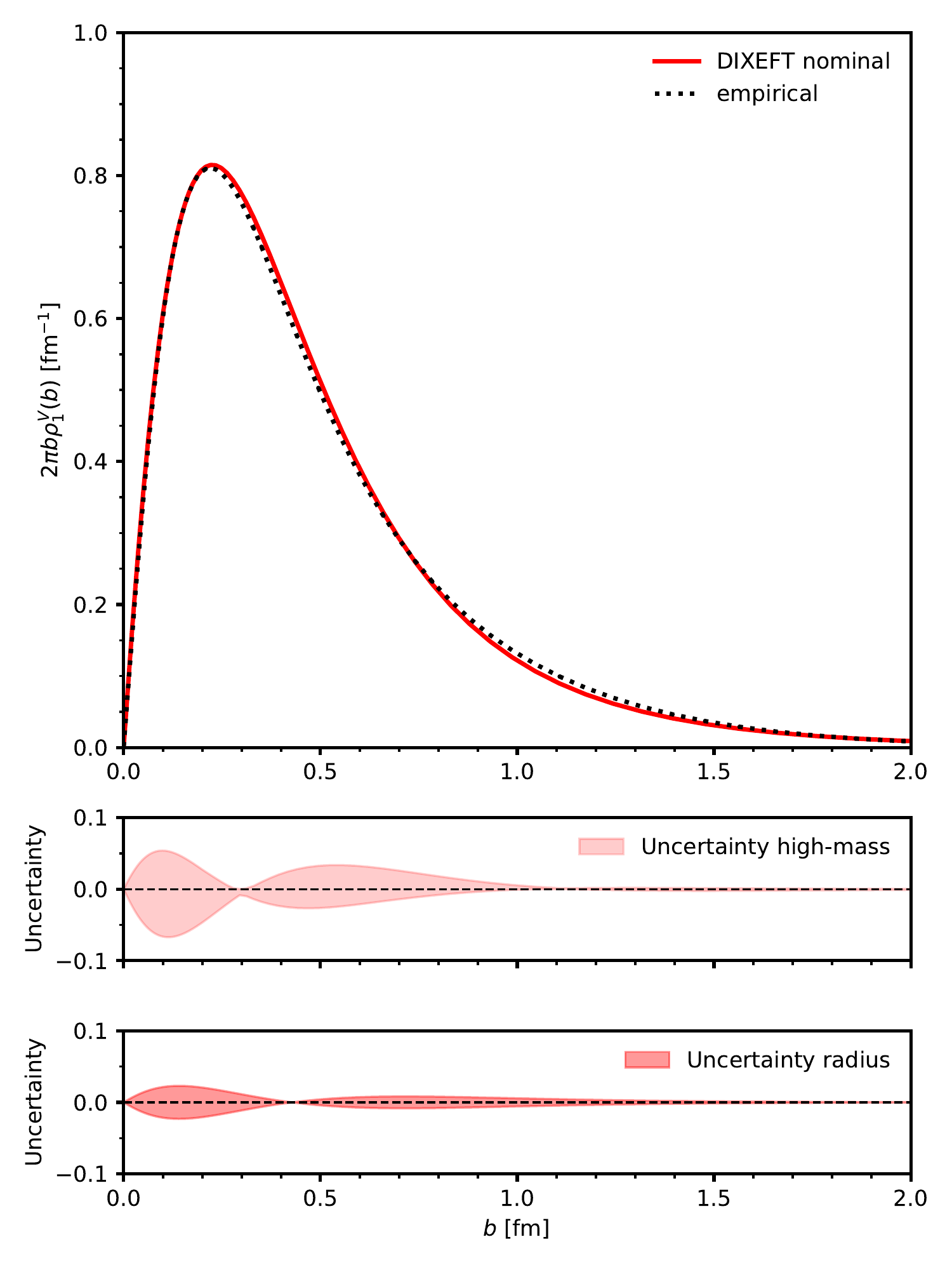} &
\includegraphics[width=.43\textwidth]{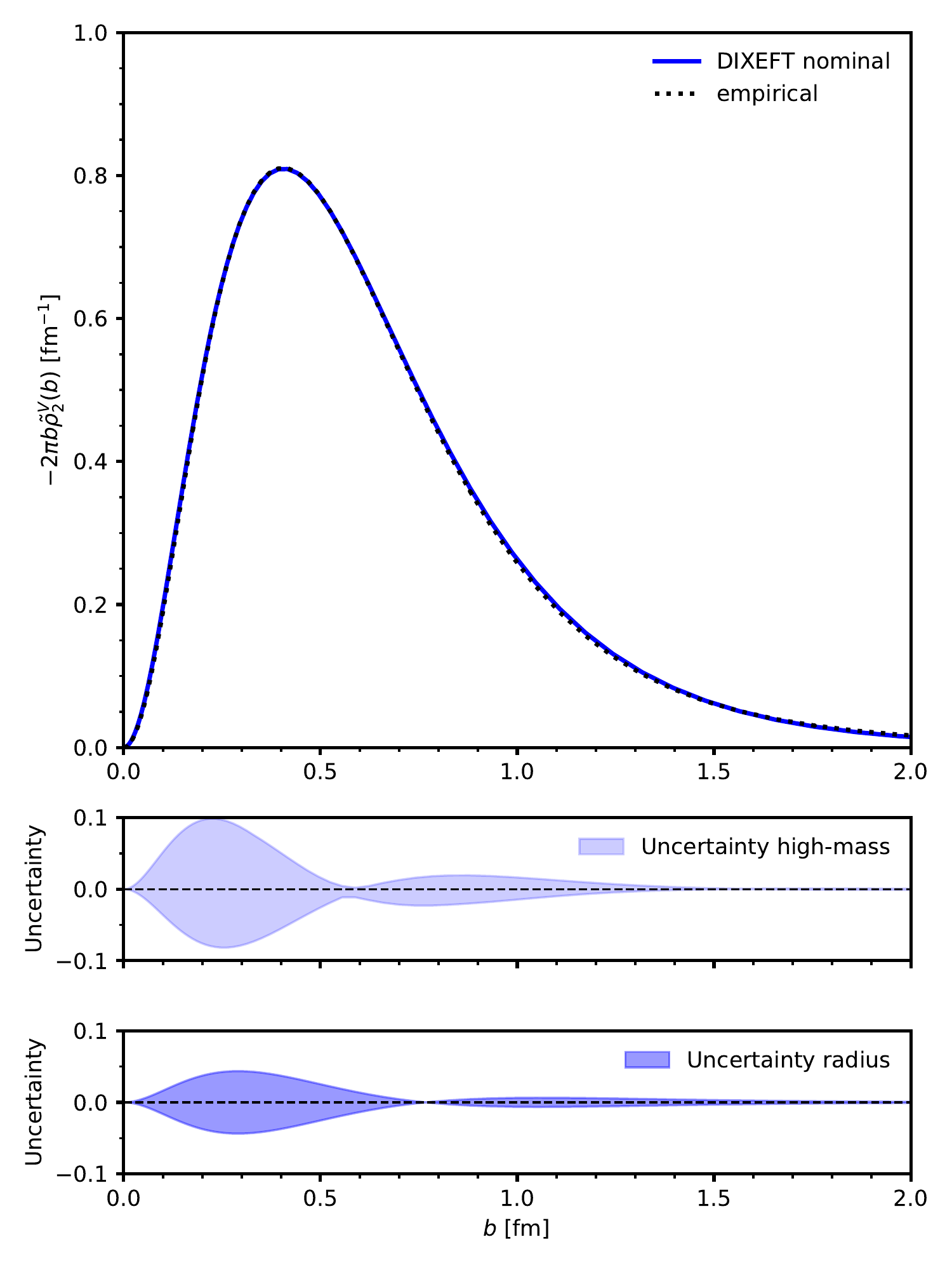}
\\[-3ex]
\end{tabular}
\caption[]{Same as Fig.~\ref{fig:rho12v_log}, but showing the radial densities
$2\pi b \rho_1^V(b)$ (left column) and $-2\pi b \widetilde{\rho}_2^V(b)$ (right column)
and their absolute uncertainties.}
\label{fig:rho12v}
\end{figure*}
Figures~\ref{fig:rho12v_log} and \ref{fig:rho12v} show our results for the isovector transverse
densities $\rho_1^V(b)$ and $\widetilde{\rho}_2^V(b)$, obtained by evaluating the dispersive integrals
Eqs.~(\ref{spectral_rho1})--(\ref{spectral_rho2t}) with the spectral functions
of Sec.~\ref{subsec:method_spectral}, and quantifying the uncertainties with
the procedure of Sec.~\ref{subsec:uncertainty}. These densities are the principal
objective of the present study. Figure~\ref{fig:rho12v_log} shows the densities $\rho_1^V(b)$ and
$-\widetilde{\rho}_2^V(b)$ on a logarithmic scale and their relative uncertainties.
Figure~\ref{fig:rho12v} shows the radial densities $2\pi b \rho_1^V(b)$ and
$-2 \pi b \widetilde{\rho}_2^V(b)$ on a linear scale and their absolute uncertainties.

One observes: (a)~The densities exhibit an exponential decrease at $b \gtrsim 0.5$ fm, as dictated
by the analytic properties of the form factor (see Sec.~\ref{subsec:dispersive_representation}).
This behavior is naturally obtained from the dispersive representation and is the principal
reason for the use of this method for the computation of peripheral densities.
(b)~The uncertainties of the densities resulting from the high-mass part of the spectral function
and from the nucleon radii show a characteristic dependence on $b$ (nodes, maxima).
This dependence is explained by the way in which these parameters influence the low-mass
and high-mass parts of the spectral functions through the dispersive sum rules
(see Sec.~\ref{subsec:method_spectral}).
(c)~The uncertainty bands are bounded at large distances and permit stable estimates of the
uncertainties of the peripheral densities. In both $\rho_1^V(b)$ and $\widetilde{\rho}_2^V(b)$,
the estimated relative uncertainties from high-mass states and radii are $\lesssim 10\%$
at distances $b > 0.5$ fm.
(d)~At $b > 0.5$ fm the relative uncertainties of $\rho_1^V$ and $\widetilde{\rho}_2^V$
are comparable. At $b < 0.5$ fm, the relative uncertainty of $\widetilde{\rho}_2^V$
is larger than that of $\rho_1^V$. This happens because at small $b$ the dispersion
integral samples the high-mass part of the spectral function, and our parametrization
of $\textrm{Im}\, F_2^V$ allows for more variation than that of $\textrm{Im}\, F_1^V$.

We emphasize that our theoretical calculations are aimed only at the ``peripheral'' densities,
with the boundary in $b$ determined by the uncertainty estimates. Our results for both
the densities and uncertainty estimates are to be understood in this sense.
At distances $b \lesssim 0.3$ fm the densities become sensitive to the details of the
high-mass states in spectral function. Our effective description is not expected to be
accurate in this region but still allows us to estimate the uncertainty and demonstrate
its increase.

Figures~\ref{fig:rho12v_log} and \ref{fig:rho12v} also show the empirical densities,
computed as Fourier transforms of the FF fit of Ref.~\cite{Ye:2017gyb}, see
Eqs.~(\ref{rho12_Fourier}) and (\ref{rho_2_tilde_def}). Because it is difficult
to quantify the uncertainties of the Fourier transform densities at large $b$
(see discussion in Sec.~\ref{subsec:results_spectral}), we show only their central value.
One observes that our theoretical results show excellent agreement with the empirical
densities at distances $b \gtrsim 0.5$ fm, where our calculations are accurate according
to our intrinsic uncertainty estimates. Our result obtained with the nominal parameters
follow the empirical densities even down to $b \rightarrow 0$. This strongly indicates
that our assumptions made in parametrizing the high-mass part of the spectral function
are realistic regarding the nominal values, and that the uncertainty of our predictions
at small $b$ could be reduced by constraining the variation of the high-mass part
through further theoretical arguments or empirical information.
\subsection{Proton and neutron densities}
%
%
\begin{figure*}[!]
\begin{tabular}{ll}
\includegraphics[width=.46\textwidth]{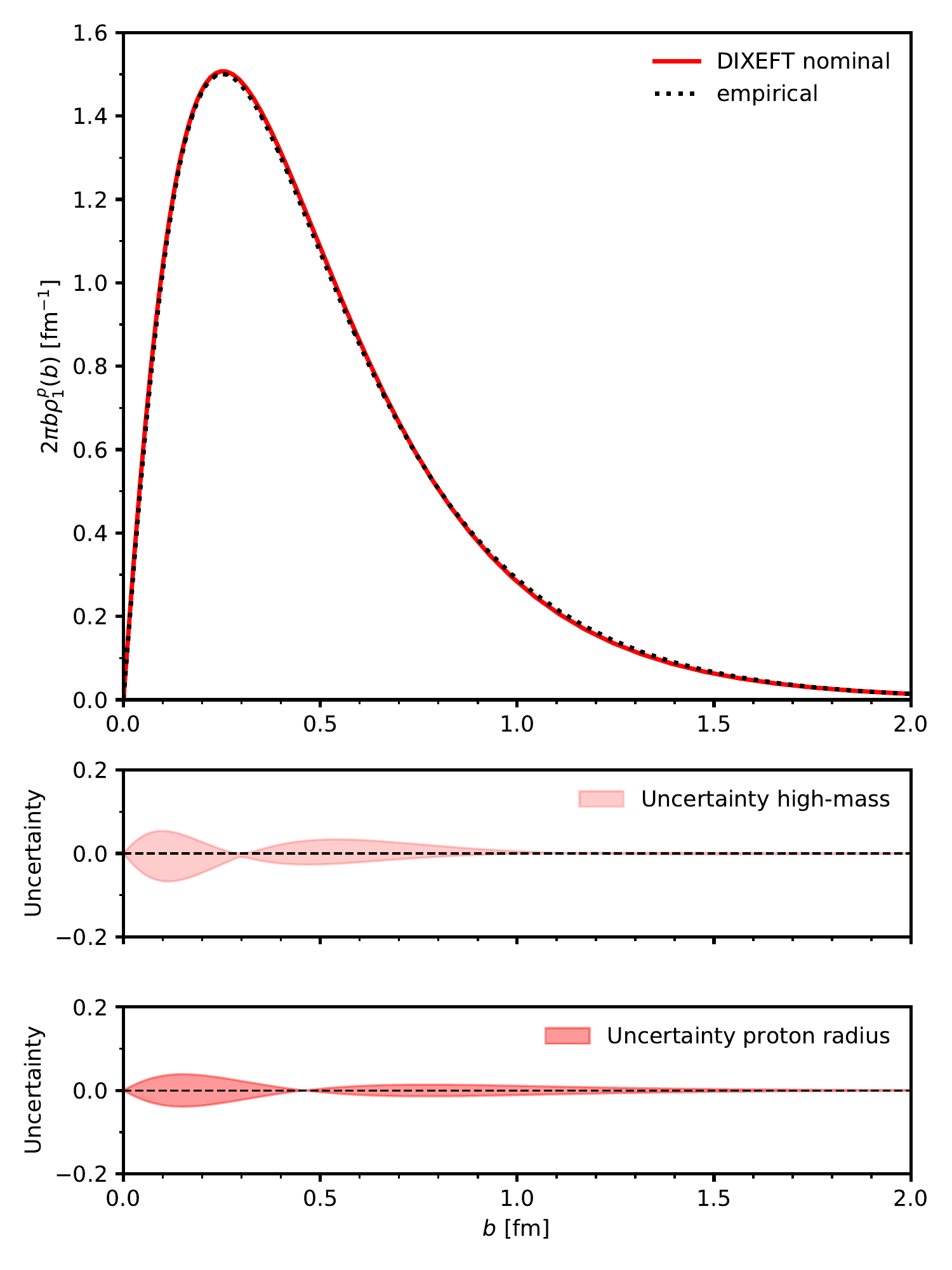} &
\includegraphics[width=.46\textwidth]{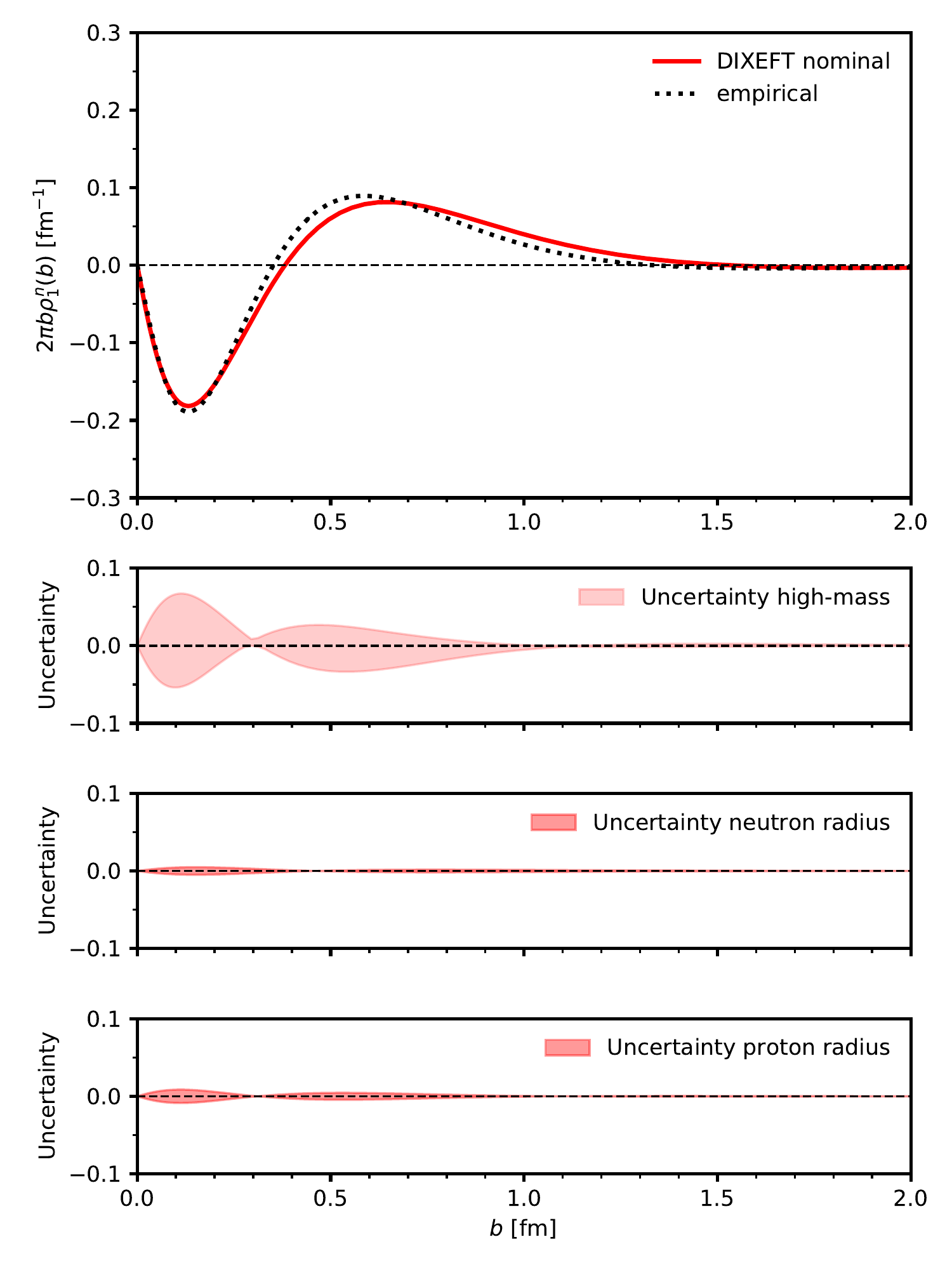}
\\[-3ex]
\end{tabular}
\caption[]{Transverse charge density $\rho_1(b)$ in the proton (left column) and neutron
(right column) obtained from the DI$\chi$EFT results (see text). In each column, the panels
show the nominal DI$\chi$EFT results and their absolute uncertainties from high-mass states
and from the nucleon radii. The empirical densities are obtained from the FF parametrizations
of Ref.~\cite{Ye:2017gyb}.}
\label{fig:rho1pn}
\end{figure*}
%
%
\begin{figure*}[!]
\begin{tabular}{ll}
\includegraphics[width=.46\textwidth]{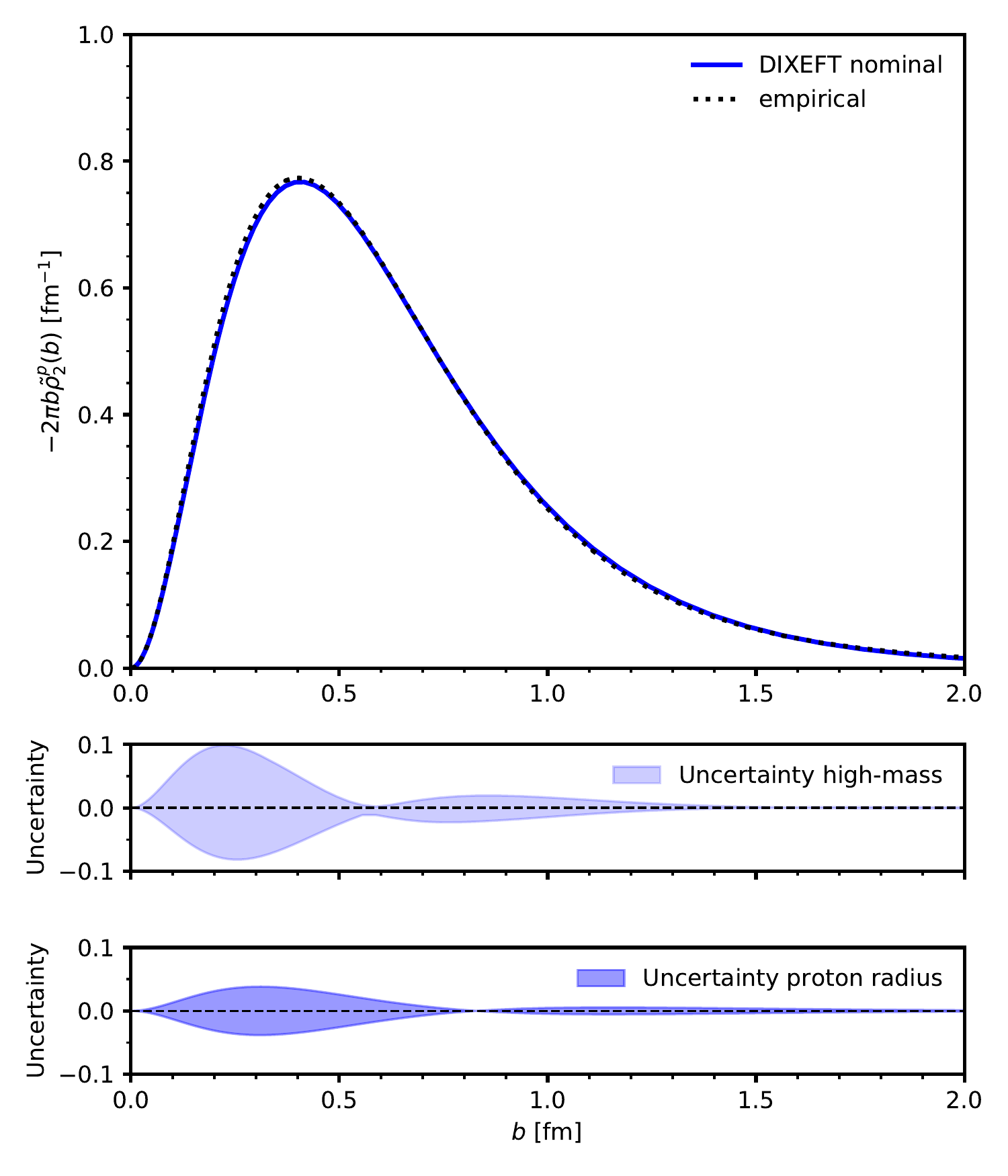} &
\includegraphics[width=.46\textwidth]{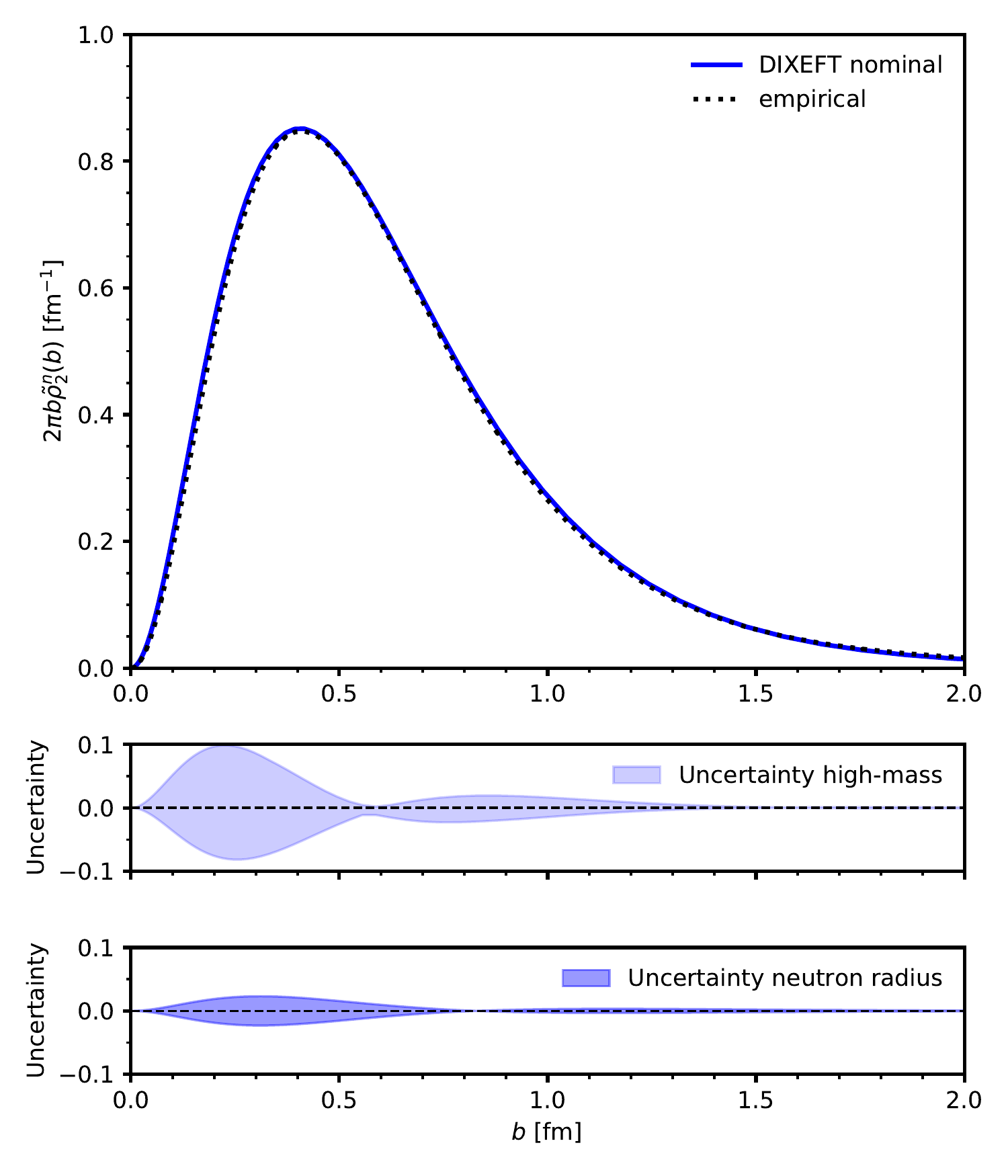} 
\\[-3ex]
\end{tabular}
\caption[]{Transverse magnetization density $\widetilde\rho_2 (b)$
in the proton (left column) and neutron (right column) obtained from the DI$\chi$EFT results (see text).
In each column, the panels show the nominal DI$\chi$EFT results and their absolute uncertainties
from high-mass states and from the nucleon radii. The empirical densities are obtained from the
FF parametrizations of Ref.~\cite{Ye:2017gyb}.}
\label{fig:rho2pn}
\end{figure*}
%
%
\begin{figure*}[!]
\includegraphics[width=.45\textwidth]{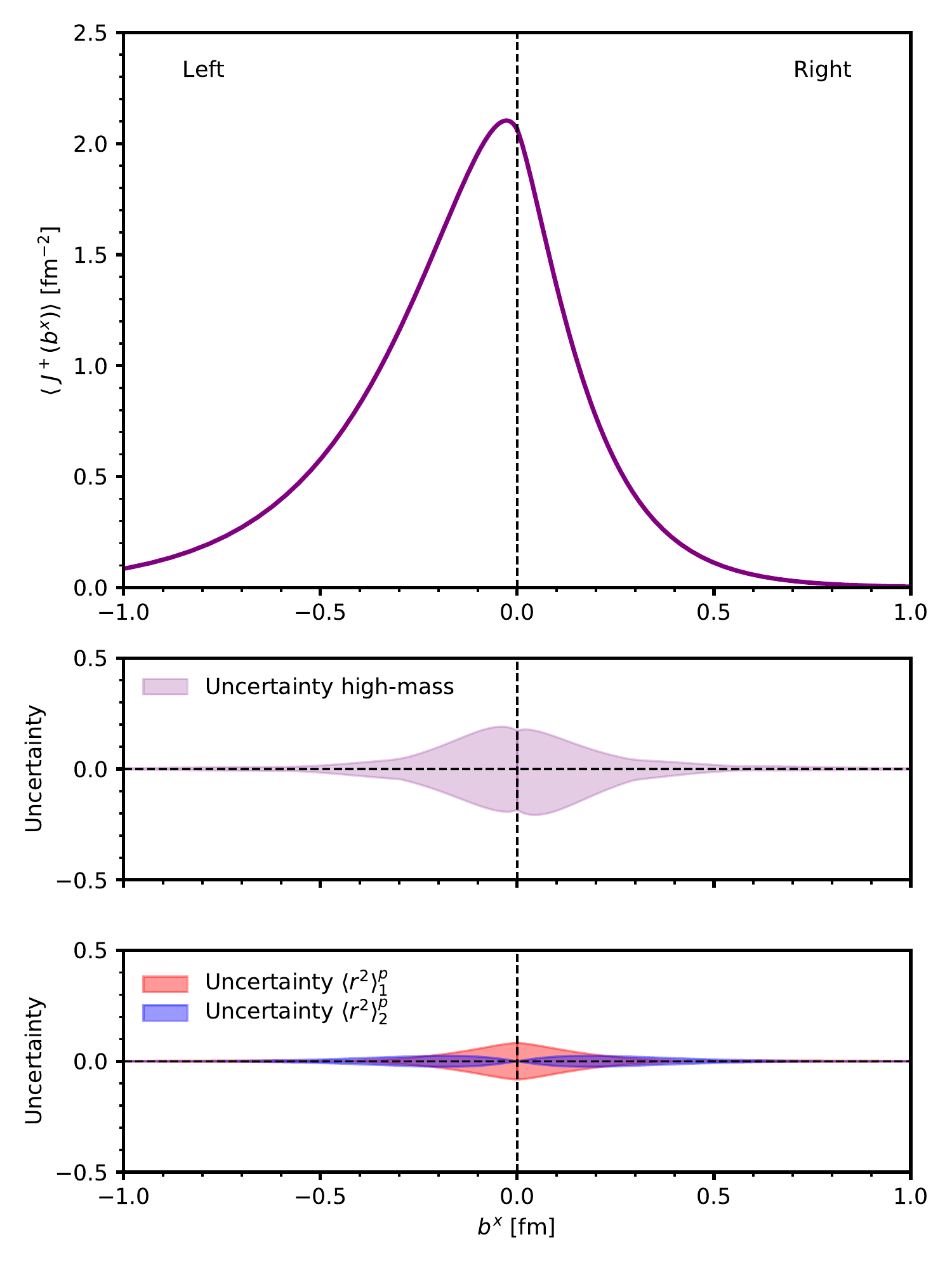}
\includegraphics[width=.45\textwidth]{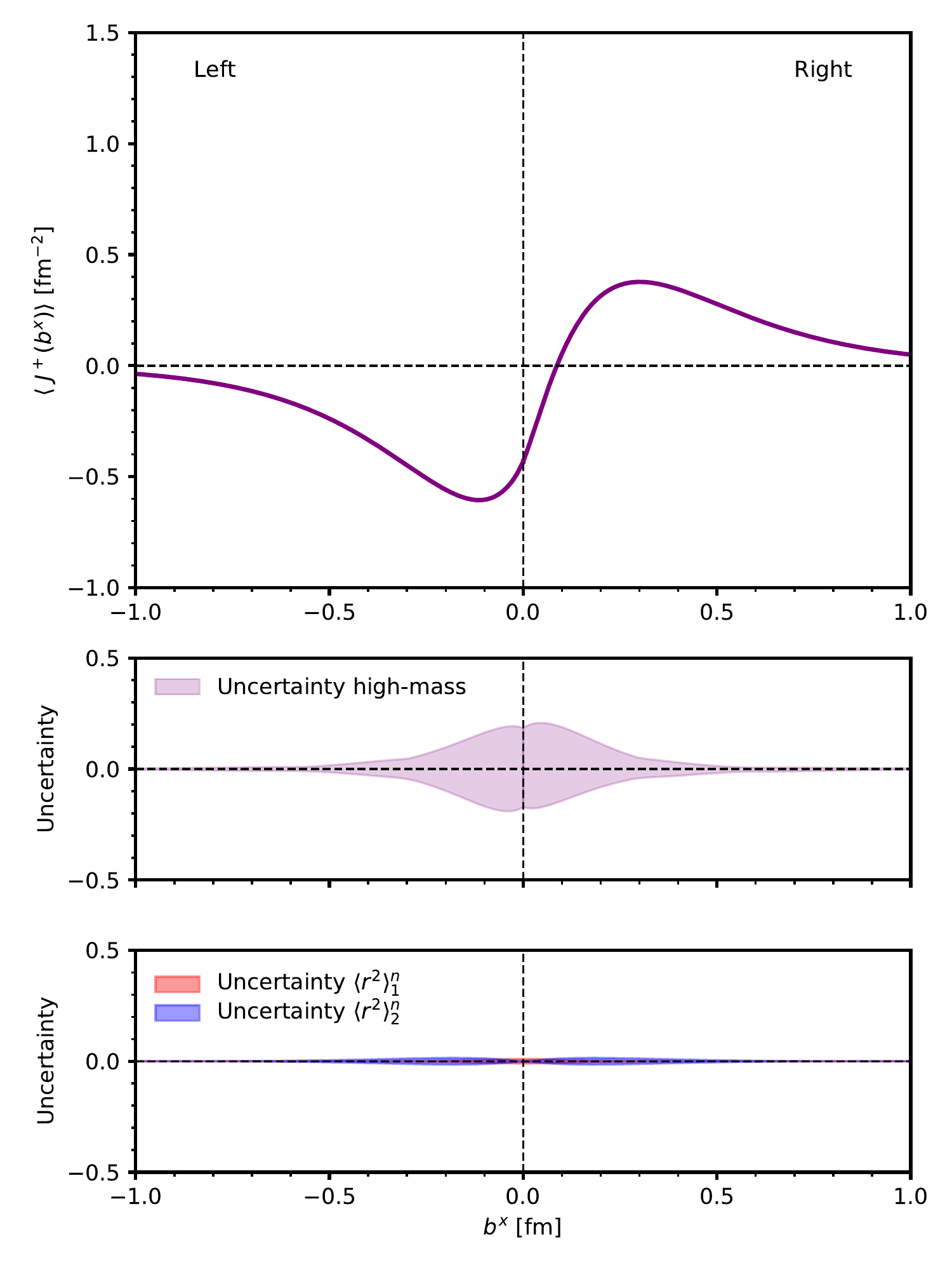}
\caption[]{The $J^+$ current density on the $x$-axis, Eq.~(\ref{rho_left_right}),
in the transversely polarized proton (left column) and neutron (right column)
with $S^y = +1/2$. In each column: {\it Upper panel:} Nominal DI$\chi$EFT result.
{\it Middle panel:} Absolute uncertainty from high-mass states.
{\it Lower panel:} Absolute uncertainty from nucleon radii.}
\label{fig:rhoLRpn}
\end{figure*}
The DI$\chi$EFT method allows us to predict the isovector densities, which contain the effect
of the two-pion states and dominate peripheral nucleon structure. In order to compute the
individual proton and neutron densities in the dispersive representation we need also the
isoscalar spectral functions. For the purposes of this study we use the parametrization
of Appendix~\ref{app:isoscalar}, which describes the isoscalar spectral function through effective
poles and implements the dispersive sum rules in a similar manner as in the isovector sector.
The parameters are fixed in terms of the isoscalar nucleon radii and their uncertainties.
The proton and neutron densities and their uncertainties are obtained by combining the
isovector and isoscalar densities. As the high-mass uncertainty of the proton and neutron
densities we take only the one resulting from the isovector densities estimated
in Sec.~\ref{subsec:isovector_densities}, which is expected to be dominant.
As the radius uncertainty of the proton and neutron densities we quote the change of
the density under variation of the nucleon's ``own'' radius, i.e., $\langle r^2 \rangle_{1, 2}^{p}$
for the proton and $\langle r^2 \rangle_{1, 2}^{n}$ for the neutron, corresponding to a
simultaneous change of the isovector and isoscalar radii. This is the dominant radius uncertainty
in most of the densities. The only exception is the neutron charge density, where the uncertainty
resulting from the change of the neutron radius is smaller than that resulting from the
proton radius, through its combined effect on the isovector and isoscalar densities.

Figures~\ref{fig:rho1pn} and \ref{fig:rho2pn} show the densities $\rho_1^{p, n}(b)$
and $\widetilde\rho_2^{p, n}(b)$ densities and their
uncertainties obtained in this way. One observes: (a)~The calculation predicts the peripheral nucleon
densities with good accuracy. In the proton charge density $\rho_1^p$, and the proton
and neutron magnetization densities $\widetilde{\rho}_2^{p, n}$,
the relative uncertainties are estimated at $\lesssim 10\%$ at $b > 0.5$ fm
(these densities are uniformly positive or negative, so that one can sensibly
quote the relative uncertainty).
(b)~In the neutron charge density $\rho_1^n$, the absolute uncertainty is estimated
at $>50\%$ near the positive maximum at $b = 0.65$ fm, and rapidly decreasing at larger $b$
(this density has different sign in different regions).
(c)~The nominal DI$\chi$EFT predictions agree well with the empirical densities at $b \gtrsim 0.5$ fm.
In particular, the theoretical calculation reproduces the behavior of the neutron density,
which changes from negative values at $b \gtrsim 1.5$ fm to positive values at
$1.5 \gtrsim b \gtrsim 0.35$ fm to negative values at $b \lesssim 0.35$ fm \cite{Miller:2007uy,Miller:2010nz}.
This behavior arises as the result of a delicate cancellation of isovector and isoscalar densities
in the different regions of $b$.
\subsection{Current densities in polarized nucleon}
Combining our results for the densities $\rho_1(b)$ and $\widetilde\rho_2(b)$, we can compute
the $J^+$ current density in the transversely polarized nucleon, Eq.~(\ref{j_plus_rho}).
In order to display the theoretical uncertainties it is useful to show a one-dimensional
projection of the two-dimensional current density. We consider the 
current density  Eq.~(\ref{j_plus_rho}) in the nucleon with $S^y = +1/2$ on the transverse $x$ axis,
where $\bm{b} = (b^x, 0)$ with $b^x <0$ or $>0$, which is given by
\begin{align}
\langle J^+ (b^x) \rangle_{\textrm{\scriptsize localized}}
\; &= \; (...) \; [ \rho_1 (|b^x|) \, + \, \textrm{sign}(b^x) \, \widetilde\rho_2 (|b^x|) ].
\label{rho_left_right}
\end{align}
This function describes the current density to the ``left'' and ``right'' when looking at the
nucleon from $z = +\infty$ (see Fig.~\ref{fig:rho12t_concept}). Notwithstanding its piecewise
definition in Eq.~(\ref{rho_left_right}), it is a smooth function of $b^x$ because
$\widetilde\rho_2(|b^x| = 0) = 0$.

Figure~\ref{fig:rhoLRpn} shows the $J^+$ current density Eq.~(\ref{rho_left_right}) in the
proton and neutron obtained from our DI$\chi$EFT results, including its theoretical uncertainty.
[The plot shows the expression in the square bracket in Eq.~(\ref{rho_left_right})
without the normalization factor denoted by (...).]
In the high-mass uncertainties we have added the uncertainty bands in $\rho_1(b)$ and
$\widetilde\rho_2(b)$ assuming no correlation between the two (the positions of the effective
high-mass poles in $\textrm{Im}\, F_1^V$ and $\textrm{Im}\, F_2^V$ are not related, and their
variation in Sec.~\ref{subsec:uncertainty} is performed independently). In the radius uncertainty
we show separately the variations of the density under the changes of the nucleon's radii
$\langle r^2 \rangle_1$ and $\langle r^2 \rangle_2$. One observes: (a)~The numerical densities
behave smoothly at $b^x = 0$, as they should. (b) The $J^+$ current densities in the proton and the neutron
exhibit a strong left-right asymmetry. In the context of the parton picture, this shows that the internal
motion due to the nucleon spin causes a significant distortion of the plus momentum distribution and attests
to the essentially relativistic character of the system. A ``mechanical'' interpretation of the left-right
asymmetry of the peripheral densities in traditional chiral EFT, as arising from the motion of a
soft pion in the nucleon's periphery, has been developed in
Refs.~\cite{Granados:2015lxa,Granados:2015rra,Granados:2016jjl,Granados:2019zjw}.
\section{Discussion}
\label{sec:discussion}
In the present work we have computed the peripheral transverse charge and magnetization densities
in the nucleon using the DI$\chi$EFT method and quantified their uncertainties. The main findings are:
(a)~The dispersive representation permits stable calculation of the peripheral densities. The
densities exhibit the exponential decrease implied by analyticity of the form factor and depend
smoothly on the parameters of the spectral function. (b)~Uncertainties can be estimated by allowing
for variation of the spectral function (functional form, parameters) and following its effect on the
densities. The procedure makes use of the particular ``information flow'' implied by analyticity
and relates the peripheral densities to the spacelike form factor in a controlled manner.
(c)~Using a minimal parametrization of the high-mass part of the isovector spectral function, the
isovector densities are computed with an estimated accuracy of $\sim \pm 10\%$ at $b \gtrsim 0.3$ fm.

In the present calculation we have not used any spacelike form factor data beyond the nucleon
radii (FF derivatives at $t = 0$) to constrain the isovector spectral functions. In particular,
we do not fit the high-mass part of the spectral function to the spacelike form factor data,
as is done is dispersive fits. [The stability condition Eq.~(\ref{stability}), controlling the
variation of the high-mass spectral function, applies to the variation relative to the nominal
theoretical prediction, not relative to the FF data.] Our results represent theoretical predictions
based on a minimal parametrization of the high-mass spectral function, and our uncertainty estimates
should be understood in this sense. It is clear that a much more accurate description could be
achieved if spacelike FF data were used to constrain the high-mass part of the isovector
spectral functions. Our estimates of the high-mass uncertainty therefore should not be regarded
as ``final,'' but rather as showing how far one can go without fitting spacelike FF data.

The methods developed here enable an EFT-based computation of the transverse densities down
to distances $b \gtrsim 0.5$ fm. At such distances the transverse densities can be described
in approaches using other effective degrees of freedom, e.g.\ quark models. This makes it
possible to match the EFT-based description with quark model predictions of the transverse
densities at ``intermediate'' distances and explore quark-hadron duality in new ways.
While quark models may not be able to accurately reproduce the absolute densities, they can
predict qualitative features such as the spin/isospin dependence and flavor decomposition
of the densities, which can lead to interesting conclusions when matched with the EFT description.
We plan to explore the use of transverse densities for quark-hadron duality studies in a separate work.

Some comments are in order regarding the interpretation of our results in terms of a ``pion cloud''
of the nucleon. It is true that the isovector densities at distances $\gtrsim$ 0.5 fm are
generated mostly by the two-pion states in the dispersive representation (see Fig.~\ref{fig:rho1v_frac}).
One might be tempted to explain this in a picture where a
bare nucleon fluctuates into a pion-nucleon state, and the peripheral structure arises from the
propagation of that pion. Such a picture is indeed obtained in traditional chiral EFT, where the
pion and nucleon are pointlike, and the peripheral densities emerge from the propagation of the
pointlike pions. However, in our unitarity-based approach the pion is not pointlike and has an
extended structure of the same range as the nucleon, as required by unitarity. The results of
our approach should therefore not be interpreted in terms of the traditional pion could picture.
The space-time interpretation of the densities in the unitarity-based approach is an interesting
question which we plan to investigate in a future study.

In the present study we have applied our unitarity-based approach to the peripheral densities
associated with the nucleon matrix element of the electromagnetic current operator. The approach
could be extended to compute the peripheral densities of other operators whose form factors
possess a two-pion cut, such as the QCD energy momentum tensor (spin-2 operator) or the
leading-twist QCD operators whose matrix elements determine the moments of the GPDs
(twist-2, spin-$n$ operators, $n \geq$ 1). This would allow one to ``deconstruct'' not only the nucleon's
electromagnetic current but also its peripheral partonic structure in terms of EFT degrees of freedom.
One difference between the electromagnetic and the generalized form factors is that for the latter
the ``radii'' (derivatives at $t = 0$) are generally not known from independent measurements,
so that one has to adjust the procedure of fixing the parameters of the spectral functions
and recruit new sources of information.
\appendix
\section{Nucleon radii}
\label{app:radii}
In this appendix we list the values of the nucleon radii used as parameters
in the DI$\chi$EFT calculation of the spectral functions.
The Dirac and Pauli radii of the proton and neutron are defined in terms of the FF derivatives
\begin{align}
\frac{dF_1^p}{dt}(0) = \frac{1}{6} \, \langle r^2 \rangle_1^p,
\hspace{2em}
\frac{dF_2^p}{dt}(0) = \frac{1}{6} \, \kappa^p \, \langle r^2 \rangle_2^p 
\end{align}
(same for $p \rightarrow n$). They are related to the conventional electric and magnetic radii by
\begin{align}
\langle r^2 \rangle_1^p \; =& \; \langle r^2 \rangle_E^p - \frac{3 \kappa^p}{2 m_N^2},
\\[1ex]
\kappa^p \langle r^2 \rangle_2^p \; =& \; - \langle r^2 \rangle_E^p + \mu^p \, \langle r^2 \rangle_M^p
+ \frac{3 \kappa^p}{2 m_N^2} 
\end{align}
(same for $p \rightarrow n$). Here $\kappa^{p, n}$ are the anomalous magnetic moments, and
$\mu^{p, n} = Q^{p, n} + \kappa^{p, n} = (2.793, -1.913)$ are the
ordinary magnetic moments of the nucleons.

We estimate the values of $\langle r^2 \rangle_{1, 2}$ and their uncertainties from
the empirical values of $\langle r^2 \rangle_{E, M}$ and their uncertainties,
neglecting correlations between the uncertainties of $\langle r^2 \rangle_E$
and $\langle r^2 \rangle_M$. We use the following numbers and sources:
$\langle r^2 \rangle_E^p = (0.7090 \pm 0.0168)$ fm$^2$ \cite{Alarcon:2020kcz},
$\langle r^2 \rangle_E^n = (-0.1161 \pm 0.0022)$ fm$^2$ \cite{Tanabashi:2018oca},
$\langle r^2 \rangle_M^p = (0.7225 \pm 0.0170)$ fm$^2$ \cite{Alarcon:2020kcz},
$\langle r^2 \rangle_M^n = (0.7465 \pm 0.0156)$ fm$^2$ \cite{Tanabashi:2018oca}.
The proton and neutron radii $\langle r^2\rangle_{1, 2}$
thus obtained are summarized in Table~\ref{table:radii}.

We also list in Table~\ref{table:radii} the isovector and isoscalar combinations
of the radii, defined by Eqs.~(\ref{sumrules_isovector_f1_values})--(\ref{sumrules_isovector_f1_values_end})
and (\ref{sumrules_isovector_f2_values})--(\ref{sumrules_isovector_f2_values_end}),
and Eqs.~(\ref{sumrules_isoscalar_f1_values})--(\ref{sumrules_isoscalar_f1_values_end})
and (\ref{sumrules_isoscalar_f2_values})--(\ref{sumrules_isoscalar_f2_values_end}),
which enter in the sum rules for the isovector and isoscalar spectral functions.
In calculating the uncertainties we neglect correlations between the uncertainties
of the proton and neutron radii. Note that the isovector/isoscalar combinations
of $\langle r^2 \rangle_2$ defined by
Eqs.~(\ref{sumrules_isovector_f2_values})--(\ref{sumrules_isovector_f2_values_end})
and (\ref{sumrules_isoscalar_f2_values})--(\ref{sumrules_isoscalar_f2_values_end}) involve
the nucleon anomalous magnetic moments and cannot directly be interpreted as nucleon radii.
%
%
\begin{table}
\begin{tabular}{l|r|r|r|}
Type & $\langle r^2 \rangle$ [fm$^2$] & $\delta \langle r^2 \rangle$ [fm$^2$] &
$\delta \langle r^2 \rangle/\langle r^2 \rangle$ \\
\hline
$\langle r^2 \rangle_1^p$ & 0.5906 & 0.0168 & 0.0285 \\
$\langle r^2 \rangle_1^n$ & 0.0102 & 0.0022 & 0.2157 \\
$\langle r^2 \rangle_2^p$ & 0.7961 & 0.0281 & 0.0353 \\
$\langle r^2 \rangle_2^n$ & 0.7518 & 0.0156 & 0.0207 \\
\hline						     
$\langle r^2 \rangle_1^V$ &  0.2902 & 0.0085 & 0.0293  \\  
$\langle r^2 \rangle_1^S$ &  0.3004 & 0.0085 & 0.0283  \\
$\langle r^2 \rangle_2^V$ &  1.4328 & 0.0293 & 0.0204  \\
$\langle r^2 \rangle_2^S$ & $-$0.0054 & 0.0293 & $-$5.3759 \\
\hline
\end{tabular}  					      
\caption[]{Nucleon Dirac and Pauli radii and their uncertainties used as input
in DI$\chi$EFT calculation.}
\label{table:radii}
\end{table}
\section{$N$ functions}
\label{app:spectral}
In this appendix we give the expressions for the $N_i(t)$ functions appearing in the
unitarity relations for the isovector spectral functions, $\textrm{Im}\, F_{1, 2}^V(t)$
Eqs.~(\ref{unitarity_real}) and (\ref{N_def}). We do not compute these functions
explicitly but express them in terms of our earlier results for the $J^1_{\pm}(t)$
functions appearing in the unitarity relations for the $\textrm{Im} \, G_{E, M}^V(t)$
spectral functions \cite{Alarcon:2017lhg}.

The Dirac/Pauli FFs $F_{1, 2}(t)$ and the electric/magnetic FFs $G_{E, M}(t)$
are related by
\begin{align}
G_E(t) &= F_1(t) + \frac{t}{4 m_N^2} F_2(t),
\label{GE_from_F}
\\[1ex]
G_M(t) &= F_1(t) + F_2(t) ,
\label{GM_from_F}
\end{align}
or, inversely,
\begin{align}
F_1(t) &= \left[ G_E(t) - \frac{t}{4 m_N^2} G_M(t) \right]
\left/ \left(1 - \frac{t}{4 m_N^2}\right) , \right.
\label{F1_from_G}
\\[1ex]
F_2(t) &= \left[ -G_E(t) + G_M(t) \phantom{\frac{0}{0}}\hspace{-.7em} \right] 
\left/ \left(1 - \frac{t}{4 m_N^2}\right) , \right.
\label{F2_from_G}
\end{align}
which hold for any complex $t$. The elastic unitarity relation for
$\textrm{Im} \, G_{E, M}(t)$, in its manifestly real form analogous
to Eq.~(\ref{unitarity_real}), is written as (here $t > 4 M_\pi^2$)
\begin{align}
\textrm{Im} \, G_E^V(t)[\pi\pi] &= \frac{k_{\rm cm}^3}{m_N \sqrt{t}} \, J_+^1(t) \, |F_{\pi}(t)|^2 ,  
\\[1ex] 
\textrm{Im} \, G_M^V(t)[\pi\pi] &= \frac{k_{\rm cm}^3}{\sqrt{2t}}    \, J_-^1(t) \, |F_{\pi}(t)|^2 ,
\end{align}
where $k_{\rm cm}$ is given in Eq.~(\ref{k_cm}). 
The relation between the functions $N_{1, 2}(t)$ and $J_\pm^1(t)$ is
\begin{align}
N_1(t) &= 
\left[ \frac{J_+^1(t)}{m_N} - \frac{t}{4 m_N^2} \, J_-^1(t) \right]
\left/ \left(1 - \frac{t}{4 m_N^2}\right) , \right.
\label{N1_from_J}
\\[1ex]
N_2(t) &= \left[ -\frac{J_+^1(t)}{m_N} + \frac{J_-^1(t)}{\sqrt{2}} \right]
\left/ \left(1 - \frac{t}{4 m_N^2}\right) . \right.
\label{N2_from_J}
\end{align}
Note that
\begin{align}
1 - \frac{t}{4 m_N^2} \; &= \; \frac{\widetilde{p}_{\rm cm}^2}{m_N^2}
\end{align}
where $\widetilde{p}_{\rm cm} = {\textstyle \sqrt{m_N^2 - t/4}}$ is the
unphysical nucleon center-of-mass momentum in the $\pi\pi \rightarrow N\bar N$ process
in the $t$-channel.

The explicit expressions for the $J_{\pm}^1(t)$ functions in chiral EFT at LO and NLO accuracy
are given in Ref.~\cite{Alarcon:2017lhg}. The expressions for the functions $N_{1, 2}(t)$
at this order can be obtained from those results using Eqs.~(\ref{N1_from_J}) and (\ref{N2_from_J}).
Note that in this procedure Eqs.~(\ref{N1_from_J}) and (\ref{N2_from_J}) are evaluated only in the
region of elastic unitarity in the two-pion channel, $t < t_{\rm max} \approx$ 1 GeV$^2$,
far away from the singularity at $t = 4 m_N^2$.

In our calculation at pN2LO accuracy we approximate the full N2LO corrections to the $N_{1, 2}(t)$
functions by rescaling the tree-level N2LO result, see Eq.~(\ref{N_lambda}). For the N2LO tree-level
result we use the simplified form
\begin{align}
N_i(t)[\textrm{N2LO-tree}] \; \propto \; t ,
\label{N_N2LO_tree}
\end{align}
the overall coefficients are irrelevant since they are absorbed by the parameters $\lambda_i$
in Eq.~(\ref{N_lambda}). Equation~(\ref{N_N2LO_tree}) differs from the exact N2LO tree-level result
[as obtained from Eqs.~(\ref{N1_from_J}) and (\ref{N2_from_J}) and the N2LO tree level result for
the $J_{\pm}^1(t)$] only by additive terms $\propto M_\pi^2$ with coefficient of order unity,
which are numerically small at $t =$ 20--50 $M_\pi^2$ where the pN2LO term Eq.~(\ref{N_lambda}) is significant.
The simplified form Eq.~(\ref{N_N2LO_tree}) ensures that each of the functions $N_{1, 2}(t)$
gets an pN2LO term with its own parameter $\lambda_{1, 2}$, which can then be fixed through
the dispersive sum rules.
\section{Isoscalar parametrization}
\label{app:isoscalar}
In this appendix we present a simple parametrization of the isoscalar spectral functions,
which is used in the dispersive calculations of the individual proton and neutron densities.
The isoscalar spectral functions are constructed along similar lines as the isovector
ones in Sec.~\ref{subsec:method_spectral}, but relying more on empirical information.

\begin{figure}[t]
\includegraphics[width=.48\textwidth]{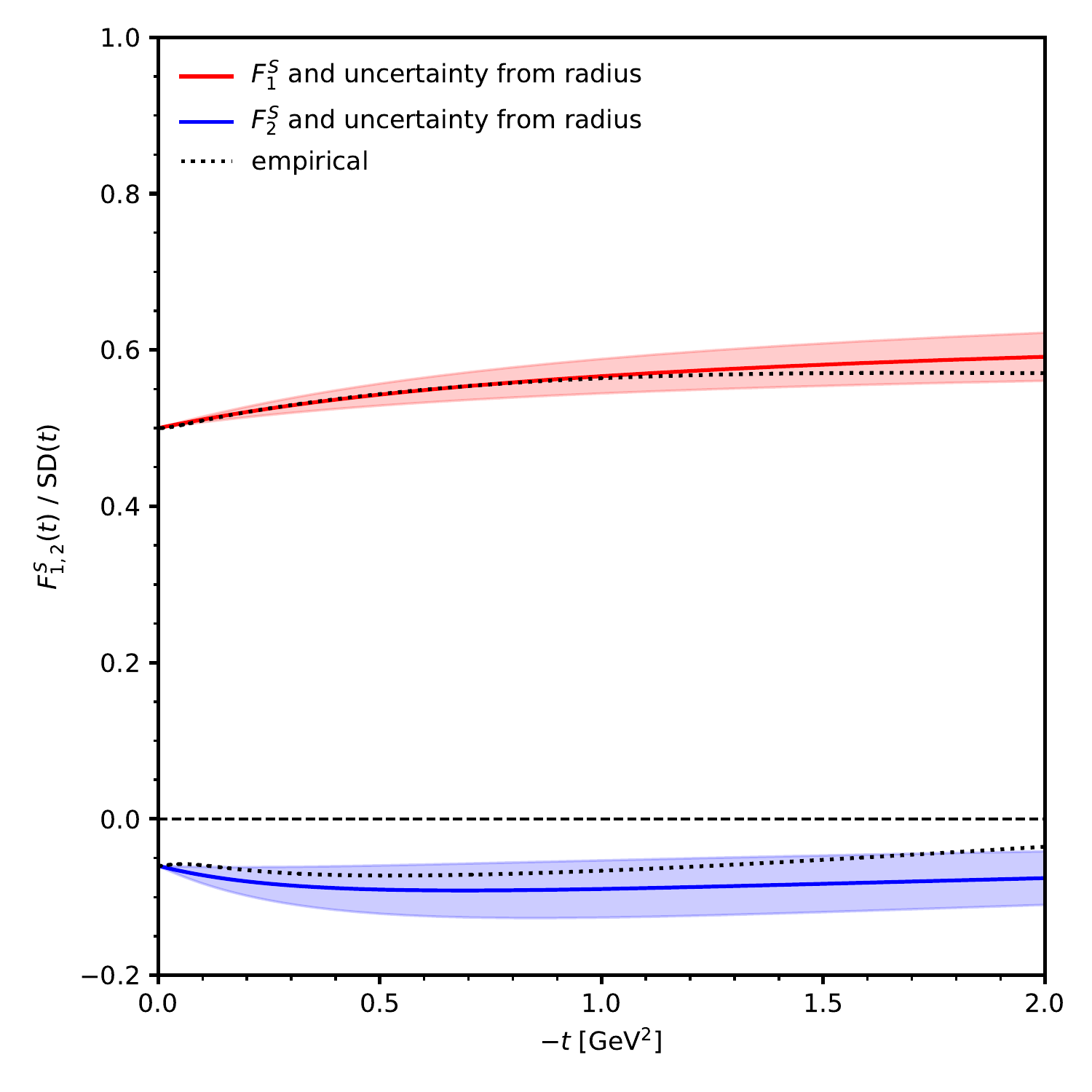}
\caption[]{Isoscalar form factors $F_1^S(t)$ and $F_2^S(t)$ obtained with our parametrization
of the isoscalar spectral functions Eqs.~(\ref{spectral_isoscalar_f1}) and
(\ref{spectral_isoscalar_f2}). {\it Solid lines:} Nominal parameters. {\it Bands:} Uncertainty resulting
from the variation of the isoscalar radii $\langle r^2 \rangle_1^S$ and $\langle r^2 \rangle_2^S$
in the range given in Table~\ref{table:radii}. {\it Dashed lines:} Empirical form factors
of Ref.~\cite{Ye:2017gyb}. All FFs are shown divided by the standard dipole FF.}
\label{fig:f12s}
\end{figure}

The isoscalar spectral function starts with the $3\pi$ channel and can be organized in a similar way
as Eq.~(\ref{spectral_isovector_pipi_highmass}) (here $i = 1, 2$),
\begin{align}
\textrm{Im}\,  F_i^S(t) =  
\textrm{Im}\,  F_i^S(t)[\pi\pi\pi] + 
\textrm{Im}\,  F_i^S(t)[\textrm{high-mass}].
\label{spectral_isoscalar_pipipi_highmass}
\end{align}
The strength in the $3\pi$ channel is overwhelmingly concentrated in the $\omega$ resonance
at $t = M_\omega^2 = 0.61$ GeV$^2$; for an estimate of non-resonant $3\pi$ contributions
in chiral EFT, see Ref.~\cite{Kaiser:2019irl}. We parametrize the $3\pi$ part of the spectral
function as ($i = 1, 2$)
\begin{align}
\textrm{Im}\, F_i^S(t)[\pi\pi\pi] &= \pi a_i^\omega \delta(t - M_\omega^2).
\end{align}
High-mass strength appears at $t \gtrsim 1$ GeV$^2$ through the $K\bar K$ channel and the
$\phi$ resonance, as well as through other hadronic states such as $\pi\rho$
\cite{Meissner:1997qt,Unal:2019eum}. We assume that the high-mass part of the isoscalar
spectral function is approximately exhausted by these states at $t \sim 1$ GeV$^2$
and parametrize the spectral functions as
\begin{align}
\label{spectral_isoscalar_f1}
&\textrm{Im}\, F_1^S(t)[\textrm{high-mass}]
\nonumber \\[1ex]
&= \pi a_1^{(S, 0)} \delta(t - t^{(S)})
\nonumber \\[1ex]
&+ \pi a_1^{(S, 1)} \delta'(t - t^{(S)})
\end{align}
and
\begin{align}
\label{spectral_isoscalar_f2}
&\textrm{Im}\, F_2^S(t)[\textrm{high-mass}]
\nonumber \\[1ex]
&= \pi a_2^{(S, 0)} \delta(t - t^{(S)})
\nonumber \\[1ex]
&+ \pi a_2^{(S, 1)} \delta'(t - t^{(S)})
\nonumber \\[1ex]
&+ \pi a_2^{(S, 2)} \delta''(t - t^{(S)}),
\end{align}
where the effective pole mass is chosen as 
\begin{align}
t^{(S)} = M_\phi^2 = 1.04 \, \textrm{GeV}^2 
\end{align}
The pole strengths $a_1^\omega, a_1^{(S, 0)}$, and $a_1^{(S, 1)}$ in $\textrm{Im}\, F_1^S$,
and $a_2^\omega, a_2^{(S, 0)}, a_2^{(S, 1)}$, and $a_2^{(S, 2)}$ in $\textrm{Im}\, F_2^S$,
are then determined by imposing the dispersive sum rules for the isoscalar
spectral functions. These sum rules are given by the formulas analogous to
Eqs.~(\ref{sumrules_isovector_f1})--(\ref{sumrules_isovector_f1_end}) and
Eqs.~(\ref{sumrules_isovector_f2})--(\ref{sumrules_isovector_f2_end})
with $V \rightarrow S$, in which the quantities on the right-hand side are now
the isoscalar charge and radius,
\begin{align}
Q^S &\equiv \frac{1}{2}(Q^p + Q^n) \; = \; \frac{1}{2},
\label{sumrules_isoscalar_f1_values}
\\[1ex]
\langle r^2 \rangle_1^S
&\equiv \frac{1}{2}
\left( \langle r^2 \rangle_1^p + \langle r^2 \rangle_1^n \right),
\label{sumrules_isoscalar_f1_values_end}
\end{align}
and the isoscalar anomalous magnetic moment and radius
\begin{align}
\kappa^S &\equiv \frac{1}{2}(\kappa^p + \kappa^n) ,
\label{sumrules_isoscalar_f2_values}
\\[1ex]
\langle r^2 \rangle_2^S
&\equiv 
\frac{1}{2}\left( \kappa^p \langle r^2 \rangle_2^p + \kappa^n \langle r^2 \rangle_2^n \right) .
\label{sumrules_isoscalar_f2_values_end}
\end{align}
The isoscalar spectral functions defined by Eq.~(\ref{spectral_isoscalar_pipipi_highmass}) et seq.\
generate remarkably accurate isoscalar nucleon form factors (see Fig.~\ref{fig:f12s})
and provide a sufficient description of the isoscalar sector for the purposes of our study. 
%
%

In the applications in Sec.~\ref{sec:results}, we take into account the uncertainty in the isoscalar
spectral function resulting from the nucleon radii. We do not attempt to assign a theoretical uncertainty
to the high-mass part of the isoscalar parametrization; this uncertainty could be estimated empirically
from dispersive fits to the data \cite{Belushkin:2006qa,Lin:2021xrc}.
\section*{Acknowledgments}
J.M.A.\ acknowledges support from the Spanish MECD grant FPA2016-77313-P.
This material is based upon work supported by the U.S.~Department of Energy, 
Office of Science, Office of Nuclear Physics under contract DE-AC05-06OR23177.
\bibliography{dixeft.bib}
\end{document}